\documentclass[preprint,12pt]{elsarticle}

\usepackage[utf8]{inputenc}
\usepackage{graphicx}   % Essential for adding images to your document.
\usepackage{subcaption}
\usepackage{xcolor}
\usepackage{float}
\usepackage{bm}
\usepackage{amsmath}
\biboptions{sort&compress,square}

\journal{arXiv}

\begin{document}

\begin{frontmatter}

\title{Probabilistic transfer learning methodology to expedite high fidelity simulation of reactive flows}

\author[inst1]{Bruno S. Soriano}
\author[inst1]{Ki Sung Jung}
\author[inst2]{Tarek Echekki}
\author[inst1]{Jacqueline H. Chen}
\author[inst1]{Mohammad Khalil}

\address[inst1]{Sandia National Laboratories, Livermore, CA 94551-0969, USA}
\address[inst2]{Department of Mechanical and Aerospace Engineering, North Carolina State University, Campus Box 7910, Raleigh 27695, NC, USA}

\begin{abstract}

Reduced order models based on the transport of a lower dimensional manifold representation of the thermochemical state, such as Principal Component (PC) transport and Machine Learning (ML) techniques, have been developed to reduce the computational cost associated with the Direct Numerical Simulations (DNS) of reactive flows. Both PC transport and ML normally require an abundance of data to exhibit sufficient predictive accuracy, which might not be available due to the prohibitive cost of DNS or experimental data acquisition. To alleviate such difficulties, similar data from an existing dataset or domain (source domain) can be used to train ML models, potentially resulting in adequate predictions in the domain of interest (target domain). However, a model based on the direct application of the knowledge gained in one domain can suffer a detrimental loss in performance when the underlying physics, and consequently the lower dimensional manifold, is dissimilar to the target domain. This study presents a novel probabilistic transfer learning (TL) framework to enhance the trust in ML models in correctly predicting the thermochemical state in a lower dimensional manifold and a sparse data setting. The framework uses Bayesian neural networks, and autoencoders, to reduce the dimensionality of the state space and diffuse the knowledge from the source to the target domain. We also investigate to what extent TL can alleviate sparsity in training data that may jeopardize the reliability of reduced-order turbulent combustion models. The new framework is applied to one-dimensional freely-propagating flame solutions under different data sparsity scenarios. The results reveal that there is an optimal amount of knowledge to be transferred, which depends on the amount of data available in the target domain and the similarity between the domains. TL can reduce the reconstruction error by one order of magnitude for cases with large sparsity. The new framework required 10 times less data for the target domain to reproduce the same error as in the abundant data scenario. Furthermore, comparisons with a state-of-the-art deterministic TL strategy show that the probabilistic method can require four times less data to achieve the same reconstruction error proving that the framework presented here can reduce the amount of data required in models based on the transport of a lower dimensional manifold.
\end{abstract}


\end{frontmatter}

\section{Introduction}

Detailed experiments and Direct Numerical Simulations (DNS) are used to understand fundamental turbulence-chemistry and molecular transport interactions in reactive flows \cite{Chen2011,Trisjono2015}. DNS complements experiments by providing more detailed information, reactive scalar concentrations, temperature and velocity,  to provide insights into a given combustion process. However, the complexity of practical fuels combined with the thin reaction layers present in realistic combustion devices results in prohibitive computational costs. These costs are driven by a large number of transported scalars and elementary reactions \cite{Lu2009} in addition to the strict spatial and temporal resolution requirements needed to resolve the multiscale and multiphysics processes governing combustion. State-of-the-art DNS  have on the order of billions of degrees of freedom (DOF) that need to be integrated over thousands of timesteps in a low-Mach solver \cite{Dalakoti2020}. The primary strategy to incorporate practical fuels in DNS is through the reduction in the number of species in the chemical mechanism models \cite{Lu2009,Savard2015,Aspden2017}. Nonetheless, mechanism reduction of complex fuels for DNS application, such as realistic gasoline surrogates and Sustainable Aviation Fuels (SAF), is often case-specific and limited to a range of thermochemical conditions. Despite the reduced number of species in the chemical mechanism, the application of DNS in reactive flows of realistic combustor geometries and conditions is often limited to a down-scaled or a canonical configuration for computational feasibility \cite{Wiseman2021,Dalakoti2020}. 

Conventional modeling strategies have been developed to expedite simulations of turbulent reactive flows. The so-called flamelet approach first proposed in Ref. \cite{Peters1984} performs a dimensionality reduction of the problem by assuming that the laminar flame structure remains unaffected by turbulence, and therefore, the one-dimensional flame structure evolution during the reaction process can be spatially tracked by a scalar that represents the combustion dynamics \cite{Peters1988}. The one-dimensional flame solutions can be pre-tabulated for a range of conditions expected to occur in the simulation. This approach has been used widely in Reynolds Averaged Navier-Stokes (RANS) and Large Eddy Simulation (LES). A similar concept has been developed to perform high-resolution simulations of turbulent premixed flames by incorporating flame stretch and curvature effects on the multi-dimensional manifold \cite{VANOIJEN2007}. The methodology has also been applied to perform high-fidelity simulations of a bluff-body premixed flame experiment at full scale \cite{Proch2017,Proch2017b}. Despite the good agreement with the experiment in terms of temperature, equivalence ratio and CO mass fraction in the radial direction, the method is valid only in the flamelet regime of the Borghi diagram with fully premixed fuel and air. Mixed-mode stratified combustion, often observed in internal combustion engines and gas turbines, as well as flames in the thin reaction zones regime, also observed in modern combustion devices, cannot be fundamentally captured by the model.

A technique based on Principal Component Analysis (PCA) \cite{Herve2010}, called Principal Component (PC) transport, has been used to expedite the DNS  of reactive flows through the systematic reduction of the composition space. The method relies on the assumption that the thermochemical process of a given combustion problem can be accurately represented by a lower dimensional manifold as presented in Ref. \cite{Mass1992}, which can be parameterized using PCs. With PC transport, the solution for the transport equations of energy, species concentrations, mass, and momentum are replaced by the solution of a significantly reduced number of PCs with mass and momentum. The method is purely data-driven so no combustion regime or mode has to be assumed \textit{a priori}. The PCs are ranked by their contribution to the amount of variance captured from the data set. As a consequence, only a subset of the total number of PCs are retained and transport equations for the PCs can be derived \cite{Sutherland2009} and transported in the physical domain. PC transport has been used to capture a `one-dimensional' turbulence simulation \cite{ECHEKKI2015,MIRGOLBABAEI2014} and multi-dimensional turbulent premixed flames \cite{Owoyele2017, Malik2021,Malik2022, Kumar2023,Abdelwahid2023,Malik2022}. Despite the successful application of PC transport in DNS, the model performance is related to the \textit{a priori} knowledge of the composition space accessed by running simpler calculations (1D or 2D) of the same flow and mixing configuration. The successful application of PC transport requires that the training data encompass all possible aerothermochemical states encountered in the DNS. The information or knowledge gathered for a given regime of operation (source domain) may not be directly applied to a different one if the lower dimensional manifold has changed enough to deteriorate the data reconstruction. Results presented in Ref. \cite{Owoyele2017} for a three-dimensional statistically-stationary planar flame showed that the dimensionality reduction has to be performed in a spatially two-dimensional data set to correctly capture multi-dimensional effects and have a good agreement with the true DNS solution. Similarly, Dalakoti et al. \cite{Dalakoti2021} also found that PC-transport using either zero-dimensional (0D) homogeneous reactor or a 1D non-premixed flamelet dataset fails to accurately represent the combustion characteristics of a 3D spatially-developing turbulent \textit{n}-dodecane jet flame. Consequently, PC transport is dependent on the data set used for dimensionality reduction, which is limited to a specific condition. %Highly transient problems can also suffer from a poor representation of the dynamics by the lower dimensional manifold depending on how the data set used for dimensionality reduction has been generated. 
The choice of data used to obtain the PCs remains an open question for the PC transport application~\cite{Owoyele2017,Kumar2023}. Moreover, the requirement of previous auxiliary data for the entire thermo-chemical state at the same conditions as the targeted simulation hinders the application of PC transport in parametric studies since it requires extensive data generation for the accurate application of the model.

PC transport is a model for the transport of a lower-dimension manifold regardless of how the manifold was obtained. The generalization can be performed by acknowledging that the transport of a given set of variables that encode a lower dimensional manifold and provide the information to obtain a low reconstruction error is the key objective of PC transport. The conceptualization of the transport of a lower dimensional manifold enables the application of Machine Learning techniques for dimensionality reduction. Neural Network architectures, such as autoencoders, are a popular platform to perform dimensionality reduction for classification, visualization, and storage of large data sets \cite{Glaws2020,Hinton2006}. The dimensionality reduction is obtained with self-supervised learning where the objective function of the neural network is to minimize the reconstruction error of the lower dimensional manifold \cite{Wang2016}. Autoencoder Neural Networks (NN) have been used in the combustion context to perform dimensionality reduction of a hydrogen flame \cite{MIRGOLBABAEI2014}. More recently, Zhang and Sankaran \cite{Zhang2022} applied autoencoders to perform dimensionality reduction of a syngas combustion system and evolve the chemical system by solving the equations of the latent variables similarly to PC transport. Even though a better representation of the lower dimensional manifold can be achieved with autoencoders, the requirement of abundant data still limits the applicability of the model since reliable data can be costly to obtain. %The model can often operate in a data sparsity setting where not all possible states are represented by the lower dimensional manifold. 

Transfer learning (TL) can be used as a possible solution to mitigate the problem related to data sparsity in machine learning models by improving the predictability in the target domain by transferring the knowledge from a different but related source \cite{Pan2010,Zhuang2021}. In the case of a manifold transport simulation, information is stored in terms of a lower dimensional latent space in the case of autoencoders or in terms of a subset of the PC modes for PC transport. As such, transfer learning could leverage the lower dimensional manifold calibrated for similar regimes of operation by informing the PC modes or the latent space with knowledge obtained from other conditions, i.e. knowledge is transferred directly in the lower dimensional manifold. Previous traditional TL methods implicitly assumed that both source and target domains are related to each other and all the knowledge was transferred \cite{Pan2010}. However, similarity between different data sets of reacting flows is not guaranteed and differences between domains can result in negative transfer learning. According to Refs. \cite{Pan2010,Weiss2016}, negative transfer learning can be defined as the situation where transferring knowledge from the source can have a negative impact on the target. As a consequence, a predictive transfer learning model should be able to quantify the amount of knowledge to be transferred between domains (tasks). Note that computing PCA for a given dataset generally provides a unique solution of the PC modes (except for the signs of the modes), such that it is not straightforward to `partially' transfer the pre-trained knowledge of the PC modes to the target domain.

The successful application of transfer learning in combustion simulations assumes that even though the lower dimensional manifold from the source domain may result in a poor prediction in a new configuration, the combination of manifolds through transfer learning can improve predictability. Knowledge gained through a similar training task is used to improve the lower dimensional manifold prediction for the target task with limited data access (data sparsity). Different transfer learning techniques have been used in the deterministic transfer learning context: penalty term; initialization with a previous solution; and a fixed number of Neural Network layers set to be constant. In Ref. \cite{MIRGOLBABAEI2014} the authors also successfully performed transfer learning in a combustion setting by applying the trained NN to a different simulation condition. The key question to be answered is whether transfer learning applied in the lower dimensional manifold can be used to improve predictions of transport manifold simulations in a data-sparse setting.

The objective of this work is to develop a novel probabilistic transfer learning framework in the Bayesian setting to leverage the pre-trained knowledge of the lower-dimensional manifold for the target domain with limited data access. In the source domain, the manifold of the original thermo-chemical state vector is obtained using a sufficient amount of training samples in a one-dimensional (1-D) freely-propagating flame solution. The knowledge from the source domain is then utilized to improve the predictability in the target domain where the number of training samples is assumed to be sparse in terms of spatial resolutions or equivalence ratios.  A linear tied-autoencoder (AE) is developed to facilitate transfer learning of the latent space. While the architecture of the linear tied AE (without biases) is identical to that of linear PCA, the pre-trained knowledge from the source domain can be included as a penalty term of the objective function of the AE model. Therefore, the amount of knowledge to be transferred from the source to the target domain can be systematically adjusted by changing the magnitude of the penalty term during the training.

The first section of this paper presented the motivation and introduction to reduced order modeling to expedite simulations of reactive flows. Section 2 provides a description of the methodology and methods used in this study, starting from PCA followed by the presentation of the autoencoders, transport of the lower dimensional manifold, Bayesian inference, and transfer learning. Section 3 describes the data generated to test the transfer learning technique and the Bayesian Neural Network setup. The last part of the paper presents the results and discussion followed by the conclusions of this study.

\section{Methodology}

The governing equations for reactive flows involve the solution of a scalar transport equation for each species in a given chemical mechanism. Assuming  Fick's law for mass diffusion, the species continuity equation is given as

\begin{equation}
    \label{eq:transp_c}
    \frac{\partial \rho Y_k}{\partial t} + \frac{\partial \rho u_i Y_k}{\partial x_i} = \frac{\partial}{\partial x_i} \left(\rho D_k \frac{\partial Y_k}{\partial x_i}\right) + \dot{\omega}_k
\end{equation}
where $Y_k$ is the mass fraction of species \textit{k}, $D_k$ is the diffusion coefficient of species \textit{k} into the mixture and $\dot{\omega}_k$ is the reaction rate of species \textit{k}. Among others, four aspects have a significant influence on making three-dimensional DNS computationally expensive: (1) the number of species, i.e. transport equations, needed to describe the oxidation of a given fuel; (2) the spatial resolution required to correctly evaluate the gradients of all reaction layers; (3) the evaluation of $\dot{\omega}_k$ which often involves hundreds and sometimes thousands of reactions; and (4) the numerical stiffness related to the orders of magnitude differences in the reaction timescales. The reduction in the number of species of a given chemical mechanism can accommodate more realistic fuels in DNS \cite{Lu2009} due mainly to its effect on (1) and (3). In three dimensions, the total number of grid cells grows non-linearly \cite{Poinsot2022} with the spatial resolution requirements and is a key factor in the cost of a DNS. Spatial resolution can be alleviated by using the Reynolds Averaged Navier-Stokes (RANS) approximation or Large Eddy Simulation (LES), where Eq.~(\ref{eq:transp_c}) is Reynolds averaged or spatially filtered. However, the averaging/filtering of the species instantaneous transport equation leads to the closure problem for the $\dot{\omega}_k$, which is one of the main focuses of combustion modeling. 

Despite the high non-linearity in the turbulence-chemistry interactions, the thermochemical state of a given combustion process often resides on a lower dimensional manifold that can be used for modeling assumptions \cite{Mass1992}. Typically, the thermochemical state of non-premixed flames is modeled in terms of the mixture fraction, and premixed flames in terms of progress variable. These two approaches are not general since they make a strong \textit{a priori} assumption related to the combustion mode to model the problem. Furthermore, the approach assumes that the combustion problem can be modeled with only a single dimension. A better dimensionality reduction can be obtained by applying PCA where the large number of correlated variables in a given data set is reduced while retaining most of the variation present in the original data \cite{Sutherland2009}. Another strategy is to employ a Neural Network autoencoder that optimizes a model for the lower dimensional manifold that minimizes the reconstruction error. As a consequence, the dimensionality reduction of the thermochemical state by either of the last two strategies does not rely on any assumptions related to the combustion mode and provides a better basis for representing the lower dimensional manifold.

\subsection{Lower dimensional manifold technique}

The transfer learning methodology proposed in this work follows the well-established PC transport modeling already used to expedite computational fluid dynamics (CFD) of reacting flows \cite{Sutherland2009,Owoyele2017,ECHEKKI2015}. The following sections present the mathematical background for the application of PC transport in combustion simulations as well as the relevance of the proposed methodology.

\subsubsection{Principal Component Analysis (PCA)}

PCA is performed to obtain a basis to represent the data by computing the PCs in the directions that best fit the data. Given a \textit{N} $\times$ \textit{P} matrix \bm{$\Theta$}, where \textit{N} number of variables and \textit{P} number of observations, and its covariance matrix \bm{$\lambda$}, the eigenvector decomposition of \bm{$\lambda$} results in $\bm{\gamma} = \bm{\Omega} ^{-1} \bm{\lambda} \bm{\Omega}$. The matrix \bm{$\Omega$} corresponds to the orthonormal eigenvectors of the covariance matrix and is the new basis to represent the PCs of the data that are computed as $\bm{\Psi} = \bm{\Omega} \bm{\Theta}$. As highlighted in Ref. \cite{Sutherland2009}, PCA provides the observations in a rotated basis with some intrinsic properties: the basis \bm{$\Omega$} is orthonormal; it optimally represents the variance in the matrix \bm{$\Theta$}; the eigenvalues \bm{$\gamma$} provide information about the eigenvectors that are of greatest importance in the data, {\it i.e.} the larger variance in the data. 

Despite the generality of PCA to obtain a lower dimension manifold representing the correlation between variables, the analysis relies on the availability of data to perform dimensionality reduction. The data used in PCA necessarily needs to be representative of the entire process under investigation, and data sparsity can lead to a poor representation of the true correlation between variables. Nonetheless, combustion models based on PCA have been successfully used to expedite reactive flow simulations by using transport equations for the PC transport. 

\subsubsection{Autoencoders}

\begin{figure}[h!]
    \centering
    \includegraphics[width=3in,keepaspectratio=true]{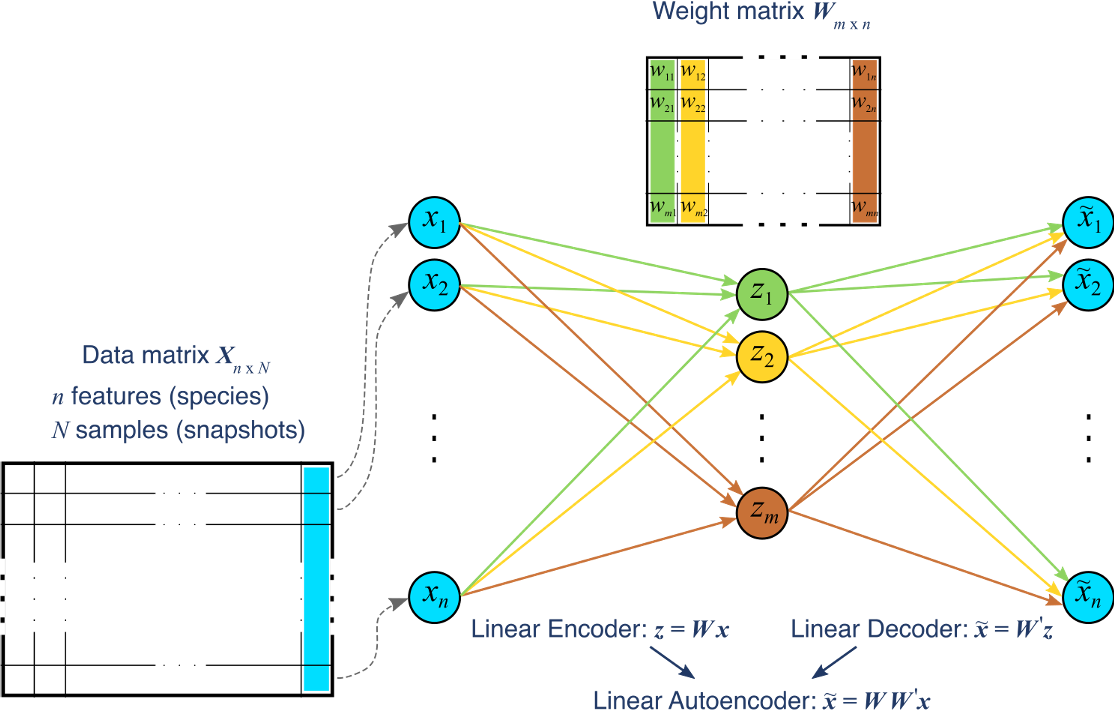}
    \caption{Schematic representation of the autoencoder architecture.}
    \label{autoencoders}
\end{figure}

Neural networks can also provide a similar dimensionality reduction with the additional benefit of leveraging third-party libraries such as TensorFlow and Pyro \cite{tensorflow2015,pyro2018}. These libraries provide the framework to perform probabilistic modeling of the lower dimensional manifold with a simpler interface for the user. Autoencoders are a popular neural network architecture to perform dimensionality reduction for classification, visualization, communication and storage of large data sets \cite{Glaws2020,Hinton2006}. It can be divided into two components as presented in Fig.~\ref{autoencoders}: the encoder part maps the variables of interest $X_i$ onto a latent space $Z_i$; and a decoder that attempts to reconstruct the original vector $X_i$ from the latent space $Z_i$ generating an output $\hat{X_i}$.

The variables $\theta_i$ (species mass fractions) pass through the hidden layer that outputs $Z_i$ following the mapping $Z_j=a\left(W_1 \theta_i + b_1\right)$, where $a$ is the activation function, $W_1$ are the weights matrix for the encoder, $b_1$ is the bias vector for the encoder, and $j<i$. The decoder part returns the data reconstruction $\hat{\theta_i}$ by performing $\hat{X_i}=a\left(W_2 Z_j + b_2\right)$, where $W_2$ are the weights for the decoder, and $b_2$ is the bias vector for the decoder. Therefore, the optimization process searches for $W_1, W_2, b_1, b_2$ that minimize the reconstruction error from $\theta_i$ and $\hat{\theta_i}$. Several hidden layers can be included in the neural network architecture for dimensionality reduction, nonetheless, only one layer is used in our model to mimic the PCA approach.

Several activation functions can be used depending on the type of application for the NN \cite{Sharma2017}. However, autoencoders with linear activation functions can be seen as an application of PCA to the input data since the latent space is the projection of the data onto a lower dimensional manifold \cite{plaut2018}. It should be noted that contrary to PCA, the lower dimensional latent space is not correlated and is not sorted by variance. Nonetheless, the framework presented here does not depend on the ortho-normality of the modes, but it relies on the correct representation of the lower dimensional manifold and the data reconstruction performed by the decoder. The NN developed in this study applies the same weights matrix for both encoder and decoder yielding the so-called ``tied autoencoders'', i.e. $W_1^T = W_2 = W$. 

\subsubsection{Transport of the lower dimensional manifold}

In the traditional PC transport, the dimensionality reduction is performed by applying PCA on the thermochemical vector obtained for the entire data set at different locations and times which can be obtained experimentally or through numerical simulations. The thermochemical vector is defined as $\bm{\theta} = (T,Y_i)^T$, where \textit{T} is the temperature and $Y_i$ is the mass fraction of species \textit{i} at a given spatial location. The principal components are related to the original thermochemical vector as

\begin{equation}
    \bm{\Psi} = \textbf{Q}^T \bm{\theta}
\end{equation}
where $\bm{\Psi}$ is the principal component (PCs) vector of size \textit{N}, \textbf{Q} is the $N \times N$ matrix of orthonormal eigenvectors, $\bm{\theta}$ is the original thermochemical vector, and \textit{N} is the total number of scalars. The PCs can be ranked by the eigenvalues according to the amount of variance captured in the data.

A large reduction in the problem size can be obtained by truncating the PCs to a subset $N_{red}$ of the most energetic ones (high eigenvalues) such that $N_{red} < N$. Such simplification is performed on the basis that PCA maximizes the variance of the data in each PC direction and also provides a ranking for the directions that capture most of the variance in the data. A large reduction in the dimensionality of the problem is expected even when more than 99\% of the variance is targeted to be captured by the model. Previous applications of PC transport have shown approximately an 80\% reduction in the number of transported PCs \cite{Mirgolbabaei2013}. In this case, a new matrix \textbf{A} $N_{red} \times N$ can be defined containing the first $N_{red}$ eigenvectors of \textbf{Q}
\begin{equation}
    \bm{\Psi}^{red} = \textbf{A}^T \bm{\theta}
\end{equation}
where the superscript \textit{red} corresponds to the reduced number of PCs used to represent the data based on the amount of variance retained by the model.

The general transport equation of the thermo-chemical scalars can be written as: 
\begin{equation}
    \rho \frac{D \bm{\theta}}{Dt} = \nabla \bm{J_{\theta}} + \bm{S_{\theta}}
\end{equation}
where $\bm{J_{\theta}}$ and $\bm{S_{\theta}}$ are the source terms for the diffusive flux and chemical production/destruction, respectively, for the thermo-chemical state $\bm{\theta}$. Similarly, Sutherland and Parente proposed a transport equation for the PCs with the form
\begin{equation}
    \rho \frac{D \bm{\Psi}}{Dt} = \nabla \bm{J_{\Psi}} + \bm{S_{\Psi}}
\end{equation}
where the diffusive flux $\bm{J_{\Psi}}$ and chemical source term $\bm{S_{\Psi}}$ correspond to the source terms for the PCs. The diffusive flux are constructed for the PCs as 

\begin{equation}
    \bm{J_{\Psi}} = \rho \bm{D_{\Psi}} \nabla \bm{\Psi}
\end{equation}
given a matrix of diffusion coefficients $\bm{D_{\Psi}}$ for the PCs, which can be obtained in terms of the original thermo-chemical diffusion matrix $\bm{D_{\phi}}$ as \cite{ECHEKKI2015}

\begin{equation}
    \bm{D_{\Psi}} = \textbf{Q}^T \bm{D_{\phi}} \textbf{Q}.
\end{equation}
Assuming a linear relationship between the PCs and the thermo-chemical vector $\bm{\theta}$, the diffusive flux and chemical source terms for the PCs can be written as

\begin{equation}
    \bm{J_{\Psi}} = \textbf{Q}^T \bm{J_{\phi}}
\end{equation}

\begin{equation}
    \bm{S_{\Psi}} = \textbf{Q}^T \bm{S_{\phi}}.
\end{equation}
Similarly, one can derive the diffusive flux and chemical source terms for the subset of retained PCs, respectively, as:  $\bm{J_{\Psi}}^{red} = \textbf{Q}^T \bm{J_{\phi}}^{red}$; $\bm{S_{\Psi}}^{red} = \textbf{Q}^T \bm{S_{\phi}}^{red}$.

As previously stated, PC transport has been successfully applied to reactive flow simulations when the data available to perform dimensionality reduction correctly represents the combustion process through the PC modes \cite{Owoyele2017,ECHEKKI2015,MIRGOLBABAEI2014,Mirgolbabaei2013}. Each PC transport simulation requires a previous auxiliary simulation with the entire thermo-chemical state at the same conditions as the targeted simulation. In order to capture multi-dimensional effects, the process often involves obtaining sufficient data samples in a two-dimensional simulation. The PC modes are then extrapolated to a three-dimensional problem. Poor representation of a premixed turbulent flame propagation was identified when the PC modes were obtained from a one-dimensional solution \cite{Owoyele2017}. However, parametric studies can provide a further understanding of the overall flame behavior at different conditions and can only be achieved by PC transport if the model is correctly calibrated for each condition, i.e. at least one auxiliary simulation is needed for each targeted case. Transfer learning can be applied to reduce the costs related to generating accurate data and increase the performance of PC transport simulations provided that, in a parametric study, the lower dimensional thermo-chemical manifold has a moderate to small variation between conditions.

\subsection{Bayesian inference}

In the context of utilizing machine learning models for scientific and engineering purposes, a physical system or component is described or approximated by a forward ML model that provides a response given a set of parameters $\theta$. The model or objective function $M$ used in this work is assumed to be linear, and a combination of the observation $d$ and a measurement error $\epsilon$, such as

\begin{equation}
\label{ML_model}
    M(x,\theta) = d + \epsilon,
\end{equation}
where \textit{x} is the thermo-chemical state, \textit{d} is the reconstruction of \textit{x} and $\epsilon$ is the reconstruction error assumed to be Gaussian and independent. $\theta$ is the collection of weights for the autoencoders here described by Bayes' rule. In other words, $M$ is the autoencoder that maps $x$ with $d$ and $\epsilon$. The process of tuning such a model involves solving the inverse problem of estimating the parameters from sparse and noisy observations. In the Bayesian framework, the solution to the inverse problem is a full probability density over the parameter vector which can be used to make probabilistic predictions. 

The proposed transfer learning methodology to be presented in Section \ref{TL_subsection} requires a statistical description of the latent variables to transfer information between different training tasks. Such statistical representation can be obtained by using Bayes' rule presented in Eq.~(\ref{Bayes_rule}), 
\begin{equation}
    \label{Bayes_rule}
    p(\theta|D) = \frac{p(D|\theta) p(\theta)}{p(D)}
\end{equation}
where a posterior distribution $p(\theta|D)$ is computed by taking into account a prior distribution (or prior knowledge) $p(\theta)$, the likelihood $p(D|\theta)$ of observing the data $D$ for a given parameter $\theta$, and the log evidence $p(D|M)$ of encountering $D$. $D$ is a collection of observations $d$.  The log evidence normalizes the posterior distribution to integrate to unity and is often ignored when sampling from the posterior.

The complexity involved in the model $M$ hinders the analytical computation of the posterior distribution. Numerical methods, such as Markov Chain Monte Carlo (MCMC) sampling and Variational Inference (VI), have been developed to evaluate the posterior. In MCMC an ergodic Markov chain on the latent variables is constructed and the posterior distribution is approximated with samples from the chain \cite{Blei2017}. Despite MCMC's capability to produce exact samples from the target distribution, what cannot be guaranteed in VI \cite{Salimans2015}, this method is computationally expensive and is not used in this study. Variational Inference corresponds to a different approach where the posterior distribution is approximated in an optimization problem. This study employs VI since it is a computationally affordable methodology to be applied to large datasets or complex models. 
 The optimization process in VI involves fitting a distribution $q(\theta)$ by selecting the model parameters that minimize the KL divergence (distance between two densities) to the posterior. The KL divergence requires information about the log evidence that is difficult to compute so the optimization process uses an alternative function called the evidence lower bound (ELBO) \cite{Blei2017}. A distribution has to be specified for $q(\theta)$ in order to perform the optimization. Typically, the complexity of the adopted distribution determines the complexity of the optimization process. In the current study, we use a mean-field approximation which assumes that the latent variables are independent from each other and modeled with a Gaussian distribution.

\subsection{Transfer learning}
\label{TL_subsection}

The objective of transfer learning is to improve model predictability by transferring knowledge from different but related sources with abundant data \cite{Zhuang2021}. Throughout this paper the source domain from which knowledge will be obtained is called ``source task'' whereas the target domain to which knowledge will be transferred to is called ``target task''. As previously stated, the methodology presented in this study aims to improve the performance of models relying on the transport of a lower dimensional manifold in the scenario of data sparsity. In the case where the manifold in the source task is exactly the same as the true manifold in the target task, the optimal solution for the minimal reconstruction error is the full transfer of the source task knowledge to the target task. Conversely, no transfer learning should be performed in the case where there is no similarity between the manifolds in the two tasks. However, in the situation where there are similarities between the two tasks, the source task information can be used to partially inform the autoencoder weights through the analysis of the likelihood function in Eq.~(\ref{Bayes_rule}). The Bayes' rule provides the mechanism to further inform the lower dimensional manifold with knowledge gained prior to the analysis of the target task. This prior knowledge is obtained by training the model on other, possibly similar, source tasks with abundant data. The limitation at this point is related to the fact that the traditional Bayes' rule is not directly applicable when the prior knowledge is obtained from other data sets because the level of similarity between tasks is not known \textit{a priori}.

The proposed transfer learning framework aims to address the shortcomings in existing methodologies by determining when to apply TL, and how much knowledge to transfer. The proposed framework extends Bayes' rule to approximate the posterior distribution with the current likelihood and prior knowledge captured by the prior distribution as
\begin{equation}
    p(\theta|D) \propto p(D|\theta) p(\theta)^{\beta}
\end{equation}\label{BTL}
where $p(\theta)$ corresponds to the prior knowledge which corresponds to the posterior distribution in the source task, and $\beta$ is a tempering parameter ranging between 0 and 1 that allows the diffusion of prior knowledge to the current posterior distribution. The $\beta$ parameter controls the amount of knowledge transferred from the prior (obtained from the source task) to the posterior of the target task. In the limiting case where $\beta = 1$, the model reverts back to the traditional Bayes' rule and is also identified as a full transfer of knowledge since the prior distribution is fully informative. On the other hand, in the limiting case with $\beta = 0$ the prior is uninformative and the posterior is equal to the likelihood, {\it i.e.} no transfer learning. $\beta$ values in the range $0 < \beta < 1$ correspond to a partial transfer of knowledge with reduced effects as $\beta$ changes from 1 to 0. 

In order to provide a simplified explanation, let us consider a simple transfer learning problem with a true Gaussian distribution and two degrees of freedom. The autoencoder weights for the source task are modeled as a Gaussian distribution N($\mu_s$,$\sigma_s$) where $\mu_s$ corresponds to the mean and $\sigma_s$ to the variance. Similarly, the weights in the target task are also modeled as a Gaussian distribution N($\mu_t$,$\sigma_t$). The posterior is written as N($\mu_p$,$\sigma_p$), where the mean and variance are evaluated as 
\begin{equation}
\label{toy_problem}
  \begin{split}
    \sigma_p &= \left[ \left(\frac{\sigma_s}{\beta} \right)^{-1}  + \sigma_t \right]^{-1} \\
    \mu_p &= \sigma_p \left[ \left(\frac{\sigma_s}{\beta} \right)^{-1} \mu_s + \mu_t \sigma_t^{-1}\right].
  \end{split}
\end{equation}
In this case, the limiting cases behave as follows:
\begin{itemize}
    \item {\bf No Transfer ($\beta = 0$):} $\sigma_p = \sigma_t$ and $\mu_p = \mu_t$. The mean and variance defining the posterior Gaussian distribution are obtained from the target task.
    \item {\bf Full Transfer ($\beta = 1$):} This is consistent with {\it standard} Bayesian updating. In cases where source task data is more abundant than that of target task, and assuming the level of noise in both data sets is equal, the posterior mean and variance of each parameter closely follow the prior mean and variance.
    \item {\bf Transfer without update ($\beta \to \infty$):} $\sigma_p = \sigma_s$ and $\mu_p = \mu_s$. The mean and variance defining the posterior Gaussian distribution are not updated with the target task data and the posterior is fully defined from the source task.
    % \item When $\beta = 1$: traditional Bayesian updating with informative prior
\end{itemize}

\begin{figure}[ht!]
\centering
\begin{subfigure}[b]{.48\textwidth}
  \centering
  \includegraphics[width=\textwidth]{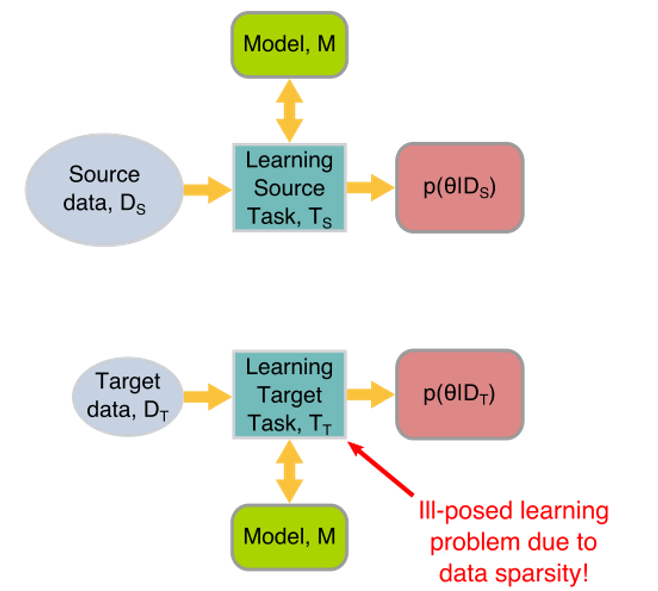}
  \caption{}
  \label{PC}
\end{subfigure}
\begin{subfigure}[b]{.48\textwidth}
  \centering
  \includegraphics[height=0.88\textwidth, keepaspectratio]{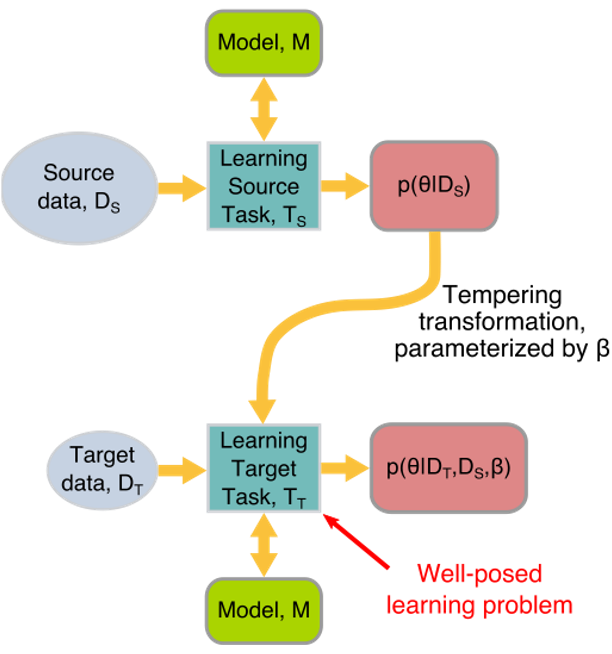}
  \caption{}
  \label{TL}
\end{subfigure}
\caption{(a) Traditional lower dimensional manifold transport model and (b) new transfer learning framework.}
\label{framework}
\end{figure}

Figure \ref{framework} presents the comparison between the traditional lower dimensional manifold transport model and the transfer learning framework proposed here. In the traditional method, the tasks are not connected and the target task is ill-posed due to data availability (sparsity) to obtain the correct lower dimensional manifold. The new methodology allows for the diffusion of knowledge from the posterior distribution of the source task ($p(\theta|D_T)$) to the target task through the prior distribution. It should be noted that the transfer of knowledge is performed for the lower dimensional manifold, {\it i.e.} autoencoder weights. In this case, the posterior distribution for the target task ($p(\theta|D_T,D_S,\beta)$) is a function of the source data $D_S$, the target data $D_T$ and the tempering parameter $\beta$.

\section{Numerical investigation}

Dimensionality reduction and transfer learning are performed using a Bayesian neural network for three different chemical mechanism sizes feasible for DNS applications. A reduced 19-species chemical mechanism for ammonia-hydrogen \cite{Jiang2020} is representative of a small composition space size. A state-of-the-art chemical model for Alcohol-to-jet fuel with 57 species \cite{Kim2021} is representative of an intermediate number of species, whereas a 98-species mechanism used to compute a gasoline surrogate multi-component mixture is representative of a large number of species. 

One-dimensional freely-propagating flame solutions were obtained in Cantera for a range of temperatures and equivalence ratios: \textit{T} = {300, 400, 450, 600} K; and $\phi=${0.5 - 2.0, in increments of 0.1}, respectively. Data sparsity was achieved in two different ways: (1) considering only a subset of equivalence ratios for a given temperature, and (2) assuming that there is a finite number of probes along the flame in order to sample the composition and temperature. The data sparsity experiment (1) reproduces the scenario when detailed data is available for a given set of conditions, {\it i.e.} DNS at one oxidizer temperature \textit{T}$_1$ and a range of equivalence ratio, and, for example, a parametric study is desired at a different oxidizer temperature \textit{T}$_2$. In this case, the transfer learning methodology developed here can improve the model predictions using only a limited number of simulations varying equivalence ratio at temperature \textit{T}$_2$. The spatial sparsity experiment (2) corresponds to a synthetic test case used to verify transfer learning robustness to capture the lower dimensional manifold and correctly reconstruct the species profiles in space.  

\begin{figure}[h!]
    \centering
    \includegraphics[width=.95\textwidth,keepaspectratio=true]{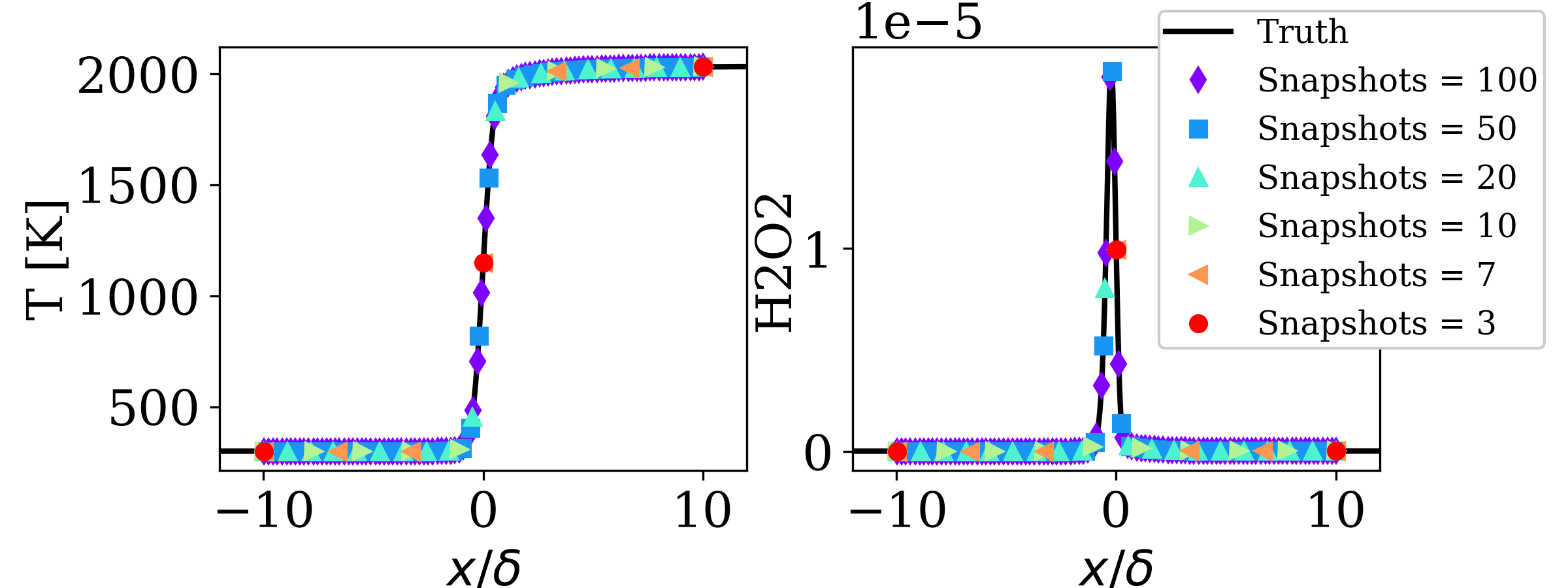}
    \caption{Spatial sparsity imposed in the fully resolved 1D freely-propagating flame solution for ammonia-H$_2$.}
    \label{fig:1D_spatial_sparsity}
\end{figure}

In order to probe the solution for approach (2), a thermal thickness is defined in terms of the temperature profile for each flame as

\begin{equation}
    \delta = \frac{T_b - Tu}{|\frac{dT}{dx}|_{max}}.
\end{equation}
where $T_b$ is the temperature in the products, $T_u$ is the temperature in the unburned mixture, and $|dT/dx|_{max}$ is the absolute value of the maximum temperature gradient.
A region encompassing 20 flame thermal thicknesses (10 ahead and 10 behind the $\frac{dT}{dx}|_{max}$ point) was linearly interpolated onto a constant spacing grid. This region is then used to obtain samples for spatial sparsity forcing. Figure \ref{fig:1D_spatial_sparsity} shows the fully-resolved temperature profile (truth) obtained with Cantera and the different equally-spaced number of samples for the spatial data sparsity cases. Even though the temperature profile can be reasonably well-captured by a small number of points across the flame brush ($\approx$ 20 points), short-lived intermediate species such as H$_2$O$_2$ may not be correctly represented for a reduced number of snapshots. Differences in how the species are captured spatially can have a direct impact on the resulting PC modes. 

The source task for the spatial sparsity tests has the reference unburned mixture conditions of $T_u$ = 300 K and $\phi$ = 1.0 and used all the data points (snapshots) in the solution for the three different chemical mechanisms used. The target tasks are sparse data points (snapshots) at equivalence ratio $\phi = 1.2$. In both cases, it is expected that the lower dimensional manifold is similar enough to enable the transfer of knowledge between tasks. 

For the sparsity in equivalence ratio, the aforementioned scenario (1), the source task is the solution of all 15 flames in the equivalence ratio range for the unburned temperature $T_u$ = 400 K. The target task corresponds to a subset of equivalence ratios ($\phi$) equally spaced in $\phi$-space for a given unburned mixture temperature. The number of flames (equivalence ratios) in the target task is varied from 10 to 1 in order to identify the effects of data sparsity in the transfer learning framework developed here.

\begin{figure}[h!]
\centering
\begin{subfigure}[b]{0.475\textwidth}
  \centering
  \includegraphics[width=\textwidth]{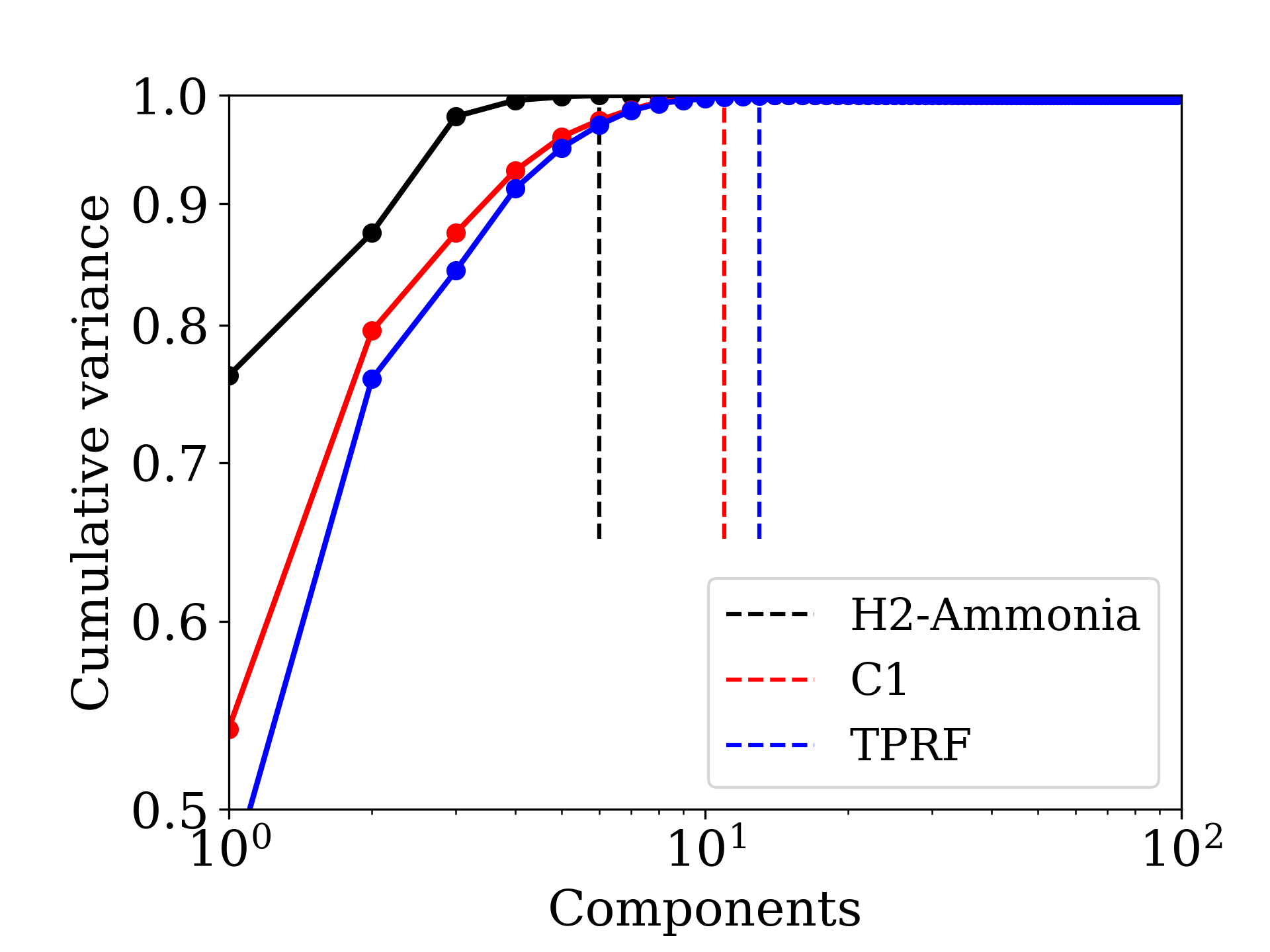}
  \caption{Spatial sparsity test}
\end{subfigure}
\hfill
\begin{subfigure}[b]{0.475\textwidth}
  \centering
  \includegraphics[width=\textwidth, keepaspectratio]{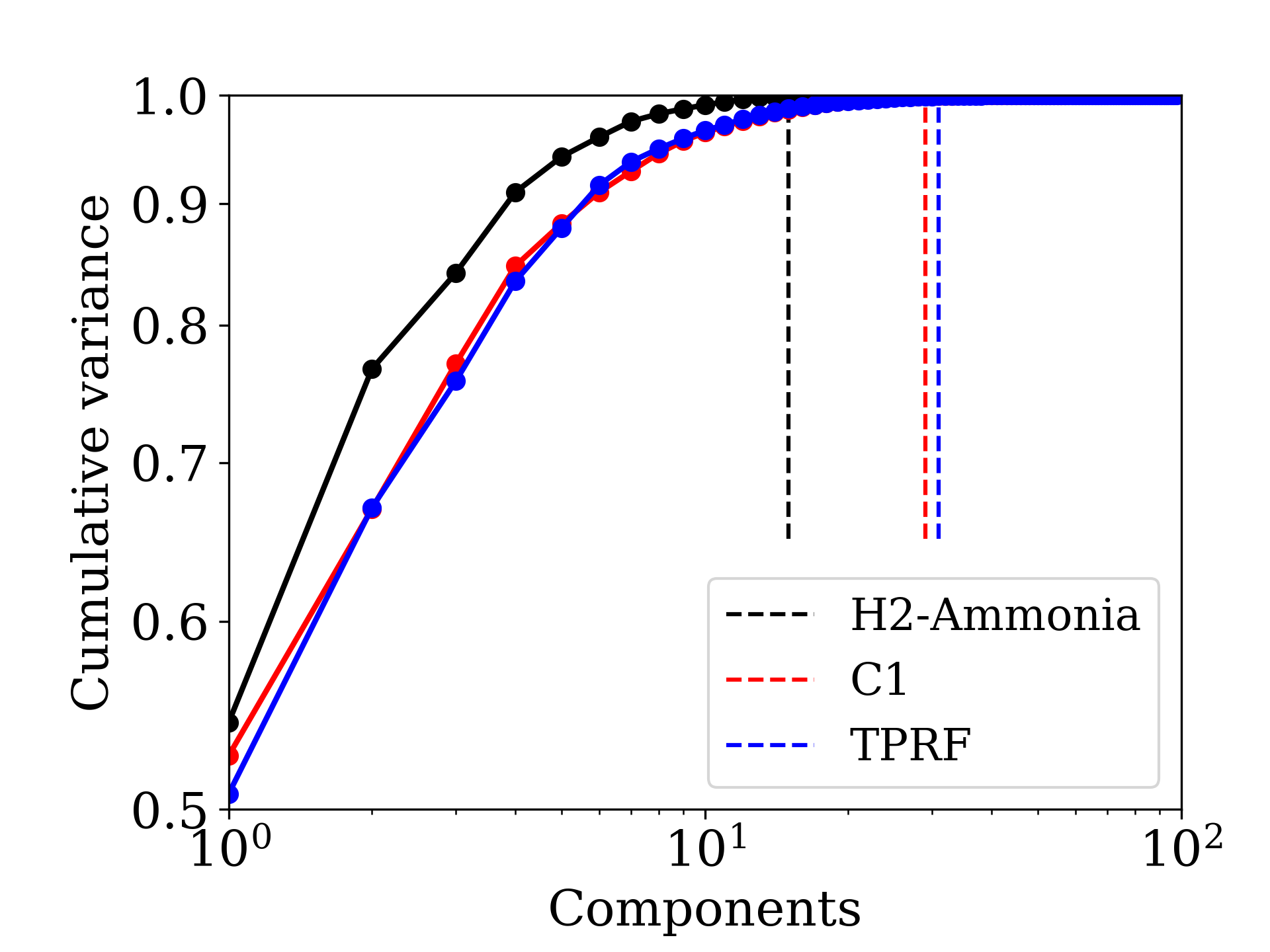}
  \caption{Equivalence ratio sparsity test}
\end{subfigure}
\caption{Cumulative variance with respect to the number of PCs retained.}
\label{fig:1D_cumulative_variance}
\end{figure}

The reduction in computational time by PC-transport is obtained through the reduction in the number of spatially-transported scalars. Only a subset of PCs are retained to capture the most of the variance (at least 99.9 \% of the variance) in the dataset. Figure \ref{fig:1D_cumulative_variance}a presents the cumulative variance captured by PCA as the number of retained components (PCs) is increased for the source task of the spatial sparsity test. The number of PCs to be retained increases with the number of degrees of freedom (number of species) in the dataset. The 19-species NH$_3$-H$_2$ flames require 6 components yielding a 68 \% reduction in the problem size, whereas the 57-species requires 11 components corresponding to an 80 \% reduction, and the TPRF flame only 13 components yielding an 86 \% reduction. It should be pointed out that the number of required components is related to the complexity of the lower dimensional manifold and does not correspond to a fixed value for a given mechanism size. The results presented in Fig.~\ref{fig:1D_cumulative_variance}b used the same methodology but for the equivalence ratio sparsity test. The number of data points (snapshots) is much larger for this study which results in a larger number of PC modes to capture 99.9 \% of the variance in the data set. The 19-species NH$_3$-H$_2$ flames require 14 components, the 57-species requires 30 components, and the TPRF flame requires 33 components.

\begin{figure}[h!]
\centering
\begin{subfigure}[b]{0.475\textwidth}
  \centering
  \includegraphics[width=\textwidth]{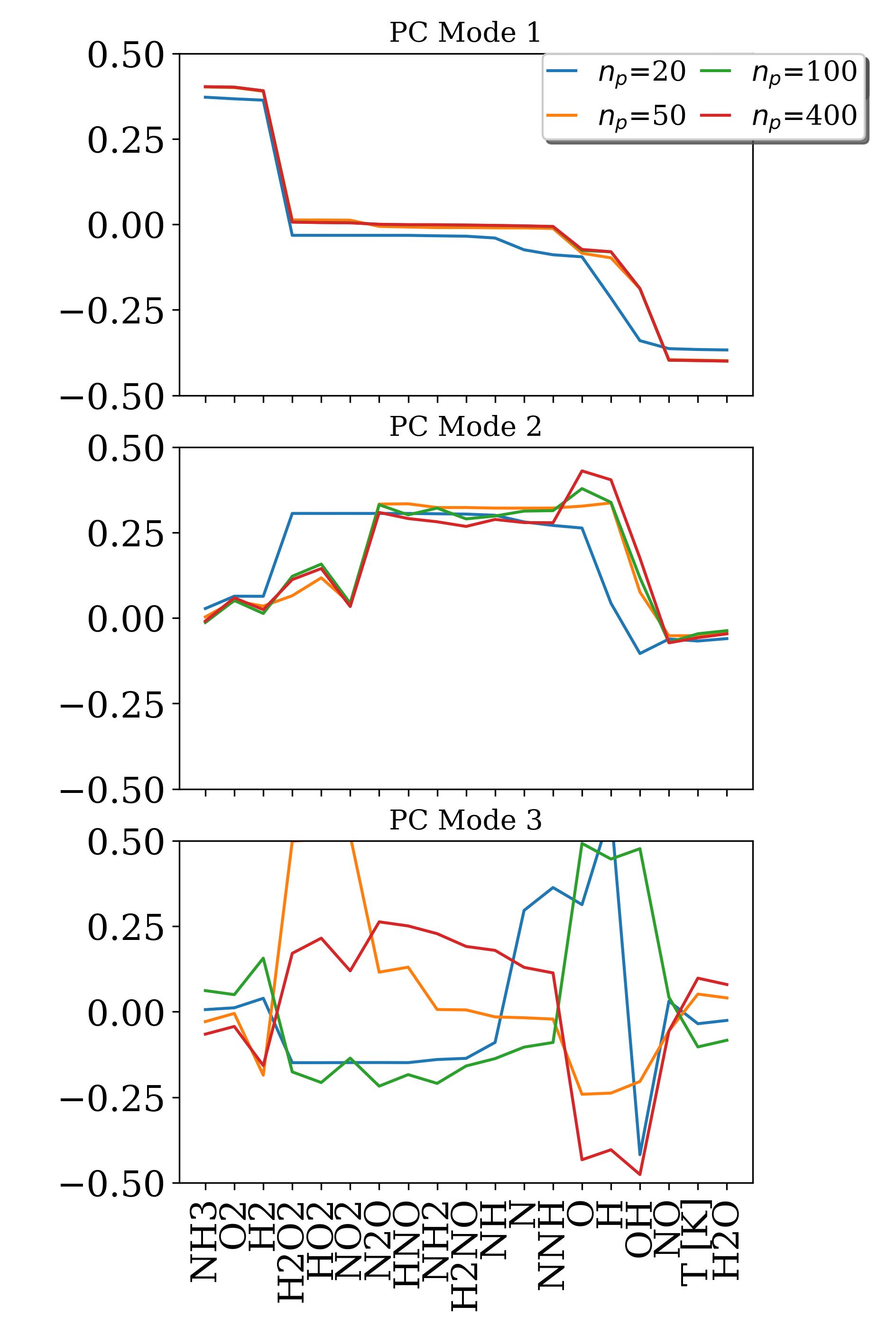}
  \caption{Spatial sparsity test}
\end{subfigure}
\hfill
\begin{subfigure}[b]{0.475\textwidth}
  \centering
  \includegraphics[width=\textwidth, keepaspectratio]{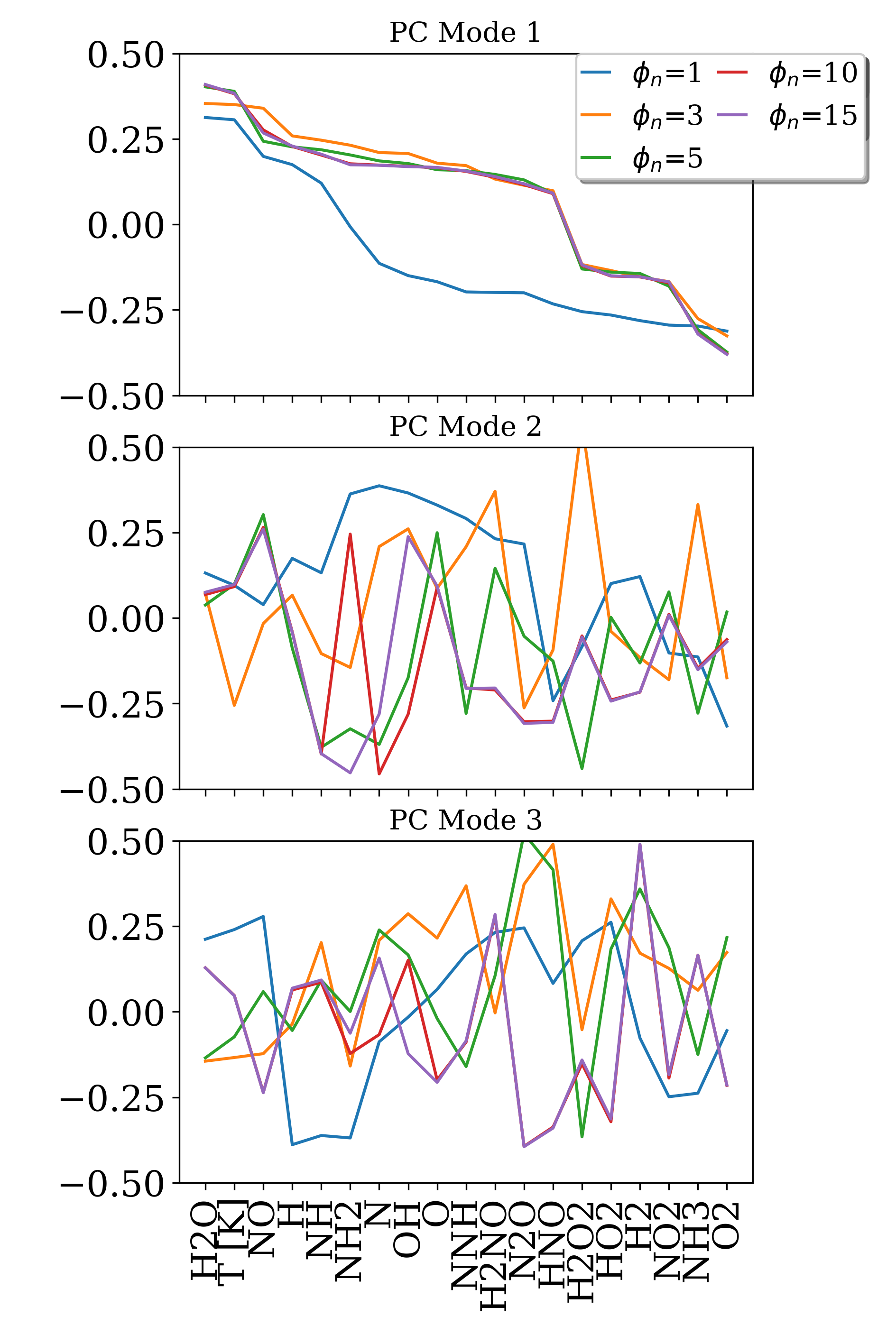}
  \caption{Equivalence ratio sparsity test}
\end{subfigure}
\caption{Distribution of the first three PC modes for the ammonia-H$_2$ mechanism.}
\label{fig:PC_modes}
\end{figure}

Transfer learning occurs through the lower dimensional manifold obtained with autoencoders. Full transfer can be performed when the PC modes are similar between source and target tasks. Conversely, no knowledge should be transferred when the tasks produce dissimilar PC modes. Partial transfer learning can improve the model performance in a sparse data setting when the tasks are sufficiently similar. Figure \ref{fig:PC_modes} presents the first three PC modes which capture more than 80 \% of the cumulative variance in the ammonia-H$_2$ chemical mechanism. The PC values are ranked in terms of the values for the first component. The figure illustrates the changes in the PC modes with data sparsity for the spatial sparsity test (Fig.~\ref{fig:PC_modes}a) and equivalence ratio sparsity test (Fig.~\ref{fig:PC_modes}b). It should be noted that the distribution of the first PC mode remains constant as the data sparsity is increased.

\subsection{Bayesian Neural Network set up}
\begin{figure}[h!]
\centering
\begin{subfigure}[b]{0.475\textwidth}
  \centering
  \includegraphics[width=\textwidth]{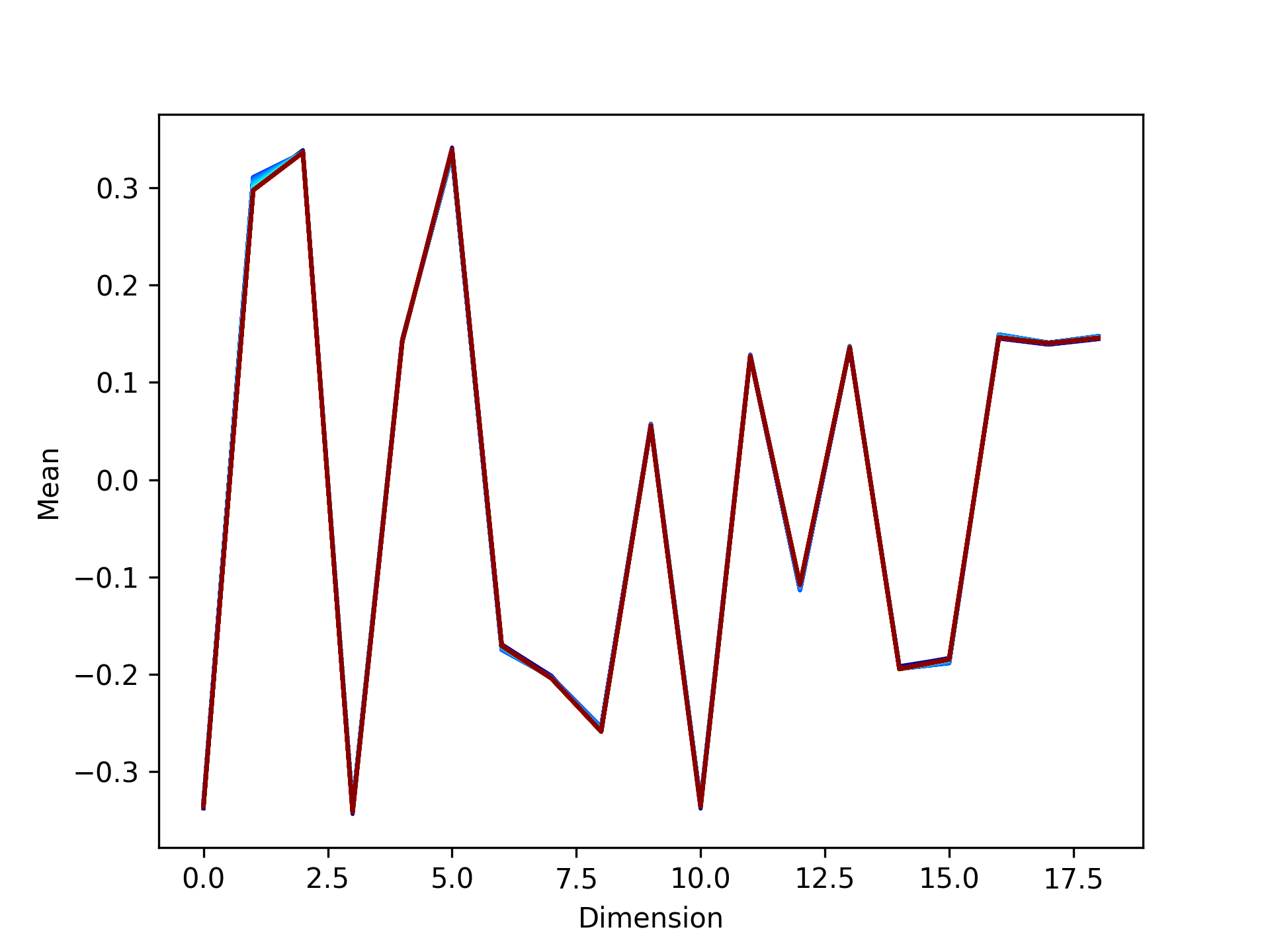}
  \caption{7 snapshots}
\end{subfigure}
\hfill
\begin{subfigure}[b]{0.475\textwidth}
  \centering
  \includegraphics[width=\textwidth, keepaspectratio]{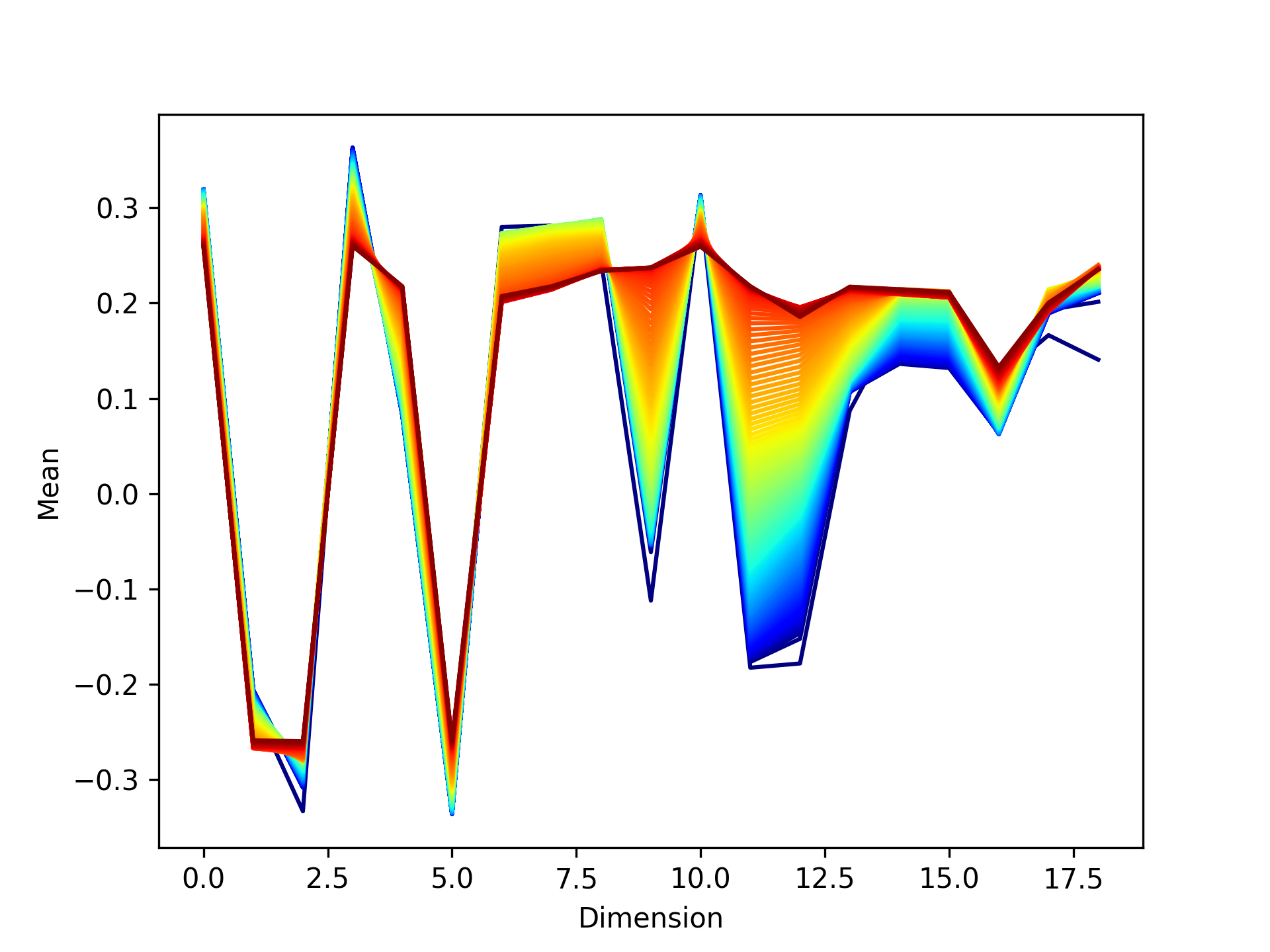}
  \caption{3 snapshots}
\end{subfigure}
\caption{Evolution of first autoencoder weights (PC mode) as a function problem dimensionality (number of scalars) during the optimization process. Evolution in epoch is colored from blue to red. Transfer learning is minimal: $\beta = 10^{-8}$}
\label{fig:PCs_evolution}
\end{figure}

The Bayesian neural network structure corresponds to a single layer of fully-connected encoder and decoder as presented in Fig.~\ref{autoencoders}. Both the encoder and decoder share the same set of weights and linear activation functions. The initial conditions for neural networks play an important role in speeding up the optimization process and finding the optimal solution. Owing to the fact that PC modes are related to autoencoder weights and that the training data set contains valuable information regarding the process to be captured, the autoencoder weights were initialized with the PC modes obtained from PCA performed with the target task. In the limiting case of large data sparsity, the PC modes should be close to the optimal solution for the autoenconders. The behavior, presented in Fig.~\ref{fig:PCs_evolution}a for the ammonia-H$_2$ flame, was verified to be true for other chemical mechanisms. 

Transfer learning was also tested in the limit where there is not enough data to perform PCA, i.e. when the number of snapshots is lower than the number of retained PCs. In this scenario, the autoencoder weights for the target task were initialized with the weights from the source task. Figure \ref{fig:PCs_evolution}b shows that the weights have a larger variation from their initial condition (in blue) up to the optimal solution (in red). The autoencoder model is optimized using the mean squared error (MSE) loss function with the Adam optimizer \cite{Adam} in conjunction with stochastic gradient descent. A parametric study was performed with respect to the number of samples used to form the ELBO estimator. A total of 400 samples was observed to be sufficient to obtain accurate results without a prohibitive increase in computational time. The learning rate for all the results presented in this work was set as $10^{-3}$ and the number of epochs was set to be a large value to ensure a converged solution.

The thermochemical state encompasses several orders of magnitude depending on the scalar; thus the model performance is related to how the data is normalized prior to dimensionality reduction. In the transfer learning context, special care should be taken for the normalization strategy in order to avoid error propagation between tasks. The normalization involves first the removal of the mean for each scalar in the thermochemical vector for the target task and then a normalization between the minimum and maximum values obtained from the source task. A different number of samples, i.e. the number of equivalence ratios or points, can result in a different mean, therefore the normalization is applied separately for each test. The strategy results in a normalized data set with zero bias and a range consistent with the source task. The data reconstruction to the original units is performed using the same minimum and maximum values from the source task, and mean from the target task.

\section{Results}

The deviation between the true solution and the predictions using the Bayesian neural network with transfer learning is presented in terms of the root mean square error normalized by the error with a unity variance (NRMSE) as

\begin{equation}
    NRMSE = \sqrt{\frac{\sum\limits_{i=1}^{n_{points}} \left(\phi_{true,i} - \phi_{p,i} \right)^2/n_{points}}{\sum\limits_{i=1}^{n_{points}} \left(\phi_{true,i} \right)^2/n_{points}}}
    \label{nrmse}
\end{equation}
where $\phi_{true,i}$ corresponds to the true solution for a given scalar at data point $i$, $n_{points}$ is the total number of points used for testing, and $\phi_{p,i}$ is the corresponding predicted quantity.

The first part of this section shows the results for the synthetic case when data sparsity is imposed by subsampling the solution spatially. The second part involves a more realistic application of transfer learning when data sparsity occurs between different flame boundary conditions.

\subsection{Transfer learning on spatially sparse freely-propagating flames}
\label{sec:spatial_sparse}

\begin{figure}[!ht]
\centering
\begin{subfigure}[b]{0.475\textwidth}
  \centering
  \includegraphics[width=\textwidth]{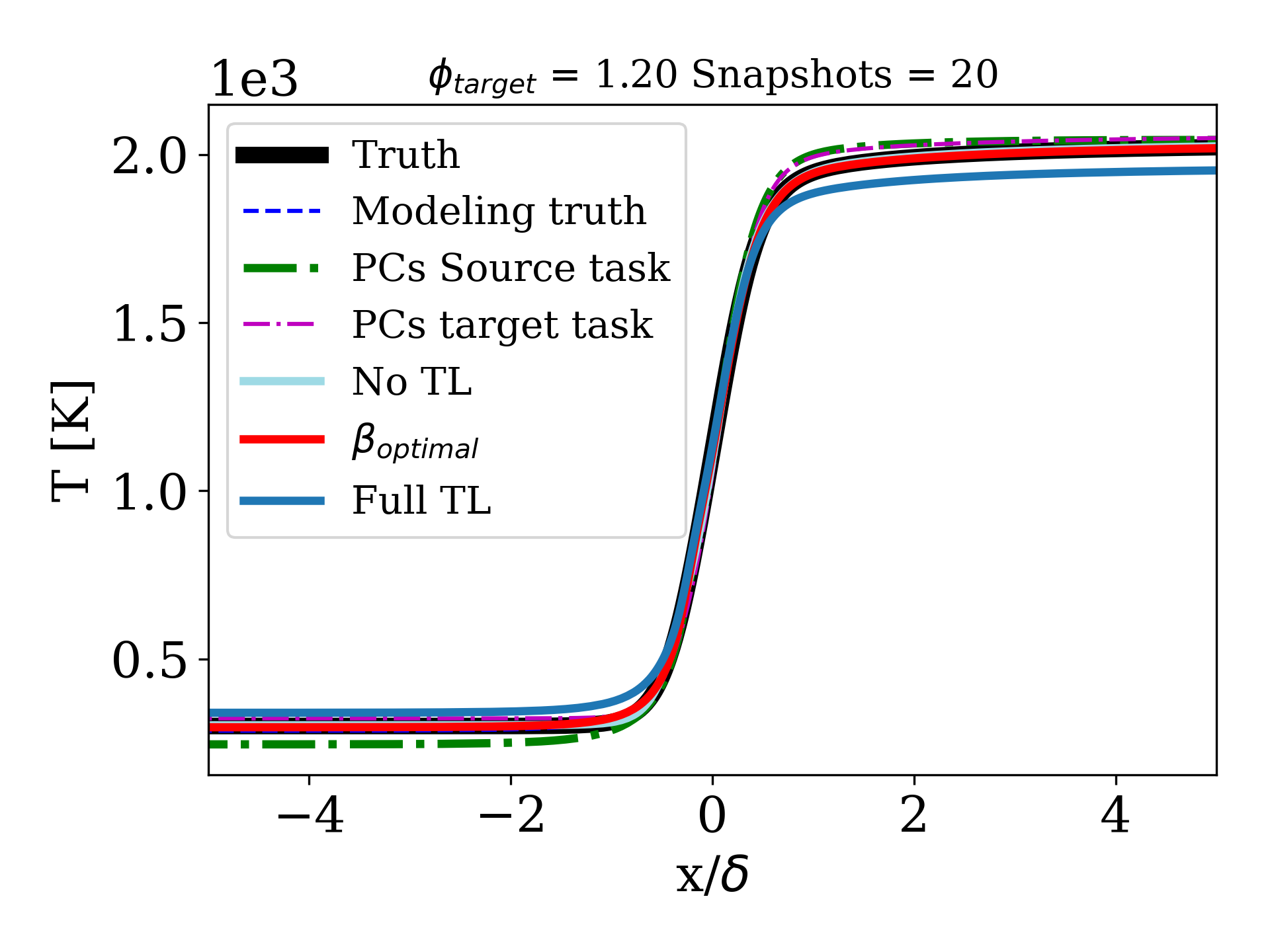}
  \caption{Ammonia-H$_2$ fuel}
\end{subfigure}
\hfill
\begin{subfigure}[b]{0.475\textwidth}
  \centering
  \includegraphics[width=\textwidth, keepaspectratio]{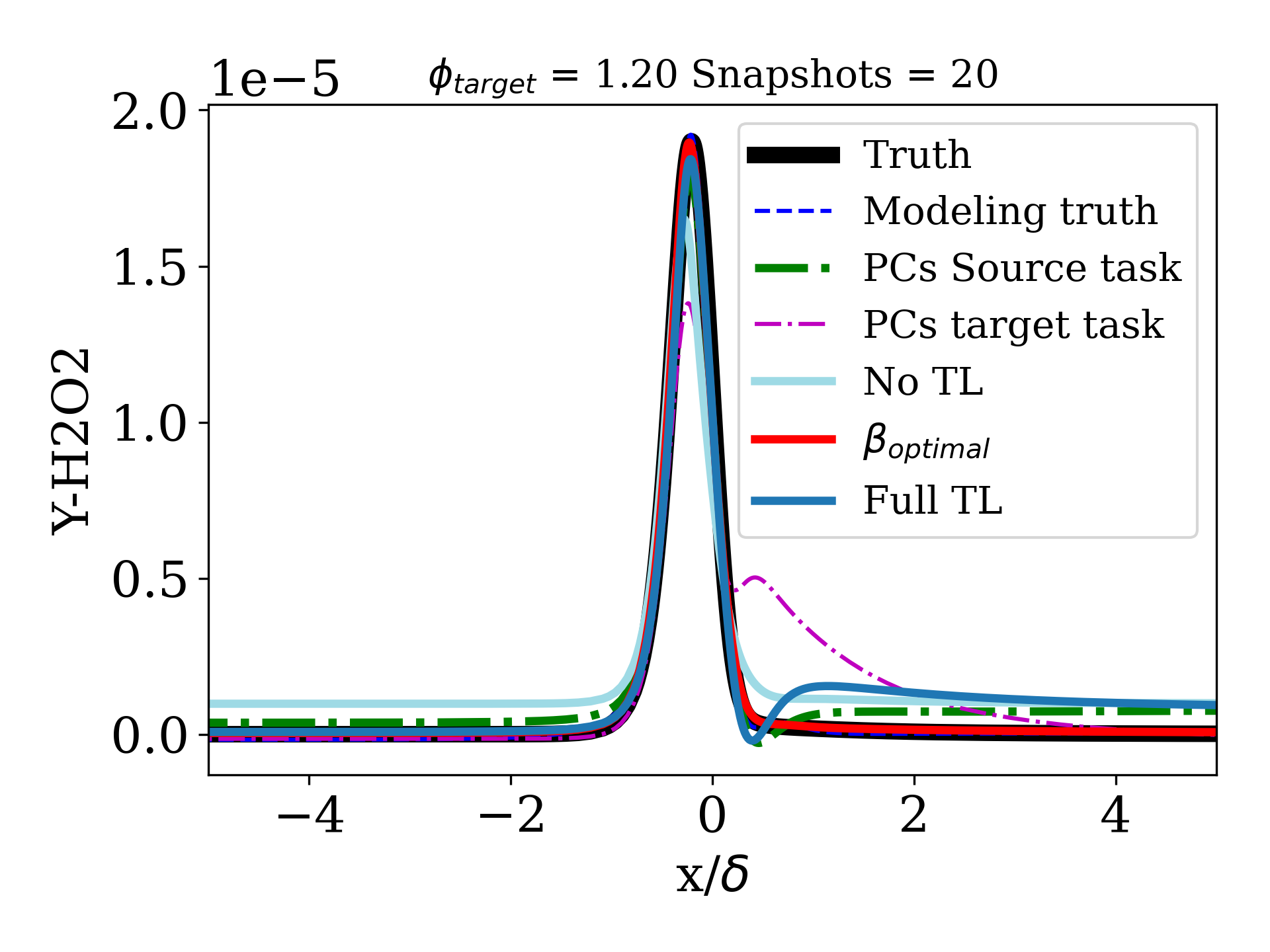}
  \caption{Ammonia-H$_2$ fuel}
\end{subfigure}

\caption{Laminar freely-propagating flame reconstruction in physical space.}
\label{fig:1D_7points}
\end{figure}

The Bayesian autoencoder is first applied to the scenario where data sparsity is imposed in a premixed flame by sub-sampling the solution in space for the target task. The stoichiometric flame results for each chemical mechanism are used as source task throughout this section, whereas the target task is a fuel-rich flame with an equivalence ratio equal to 1.2. Figure \ref{fig:1D_7points} presents the reconstruction in physical space of temperature profiles and a chemical species produced and consumed during the fuel oxidation process. Short-lived species present a challenge in the data reconstruction due to the lack of resolution to correctly capture its correlation with the other species in the dimensionality reduction process. The line labeled as ``truth'' corresponds to the 1D flame solution using Cantera, the ``modeling truth'' is the reconstruction using the retained PCs for all snapshots in the target task, and the ``PCs source task'' is the prediction when the PCs from the source task are applied to reconstruct the target data set. This corresponds to the limiting case where full transfer learning is applied and the PCs are obtained directly from PCA, i.e. no Bayesian neural network is used to capture the lower dimensional manifold. ``PCs target task'' is the reconstruction using the PCs obtained with the sparse target task snapshots. Three additional lines with different amounts of knowledge transferred are included in Fig.~\ref{fig:1D_7points}: ``No TL'' corresponds to the solution of the dimensionality reduction obtained with the BNN with $\beta \to 0$; ``Full TL'' corresponds to the case where $\beta$ = 1; and ``$\beta_{optimal}$'' corresponds to the case where $\beta$ results in the lowest reconstruction error.

The first feature to be noticed in Fig.~\ref{fig:1D_7points}a is that ``source task'' knowledge cannot be directly applied to the reconstruction of the temperature profile for the ammonia-H$_2$ flame. The unburned mixture temperature is under-predicted and the burned temperature is over-predicted. The behavior is in line with the expectations given that the flame for the source task is stoichiometric (higher adiabatic temperature) and the flame for the target task is fuel-rich (lower adiabatic temperature). Similarly, the full transfer learning strategy with $\beta$ = 1 and the PCs from the target task also result in deviations from the true temperature profile. The predictions ``No TL'' and $\beta_{optimal}$ have a good agreement with the true profile. Temperature presents a high absolute value for the first PC mode that remains unchanged for different numbers of data points used for dimensionality reduction (see Fig.~\ref{fig:PC_modes}a), therefore it is expected a good agreement with the flame solution for a sparse setting without transfer learning. Conversely, the H$_2$O$_2$ species is captured by the higher PC ranks that are more affected by data sparsity. Deviations in the oxidizer and products side as well as in the peak of H$_2$O$_2$ are observed for the limiting cases with ``PCs source task'', ``PCs target task'', ``Full TL'', and ``No TL''. Therefore, an accurate construction of the H$_2$O$_2$ profile can only be obtained with a partial transfer of knowledge where $0 < \beta < 1$.

\begin{figure}[!ht]
\centering
\begin{subfigure}[b]{0.475\textwidth}
  \centering
  \includegraphics[width=\textwidth]{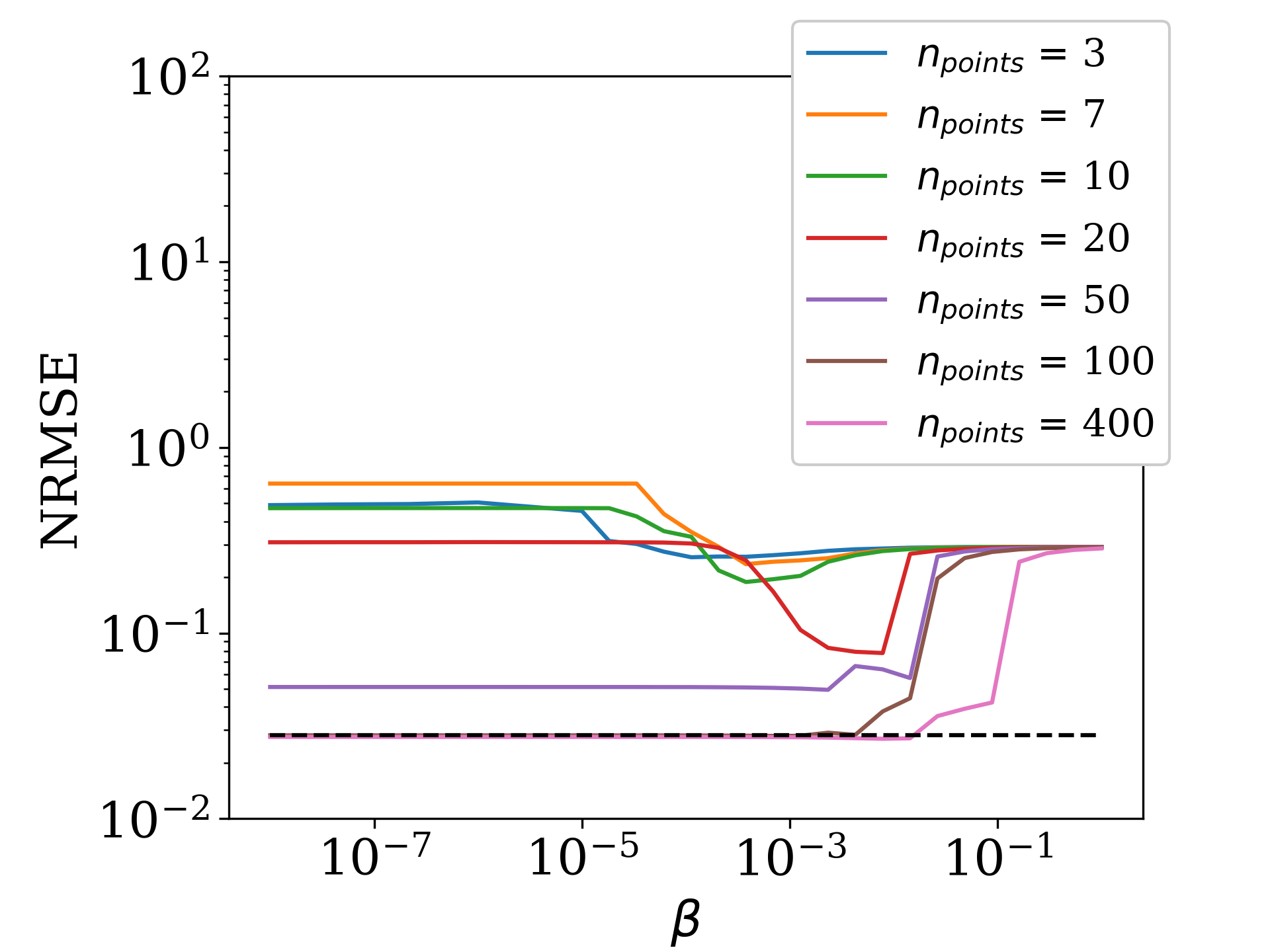}
  \caption{Ammonia-H$_2$ fuel}
\end{subfigure}
\hfill
\begin{subfigure}[b]{0.475\textwidth}
  \centering
  \includegraphics[width=\textwidth, keepaspectratio]{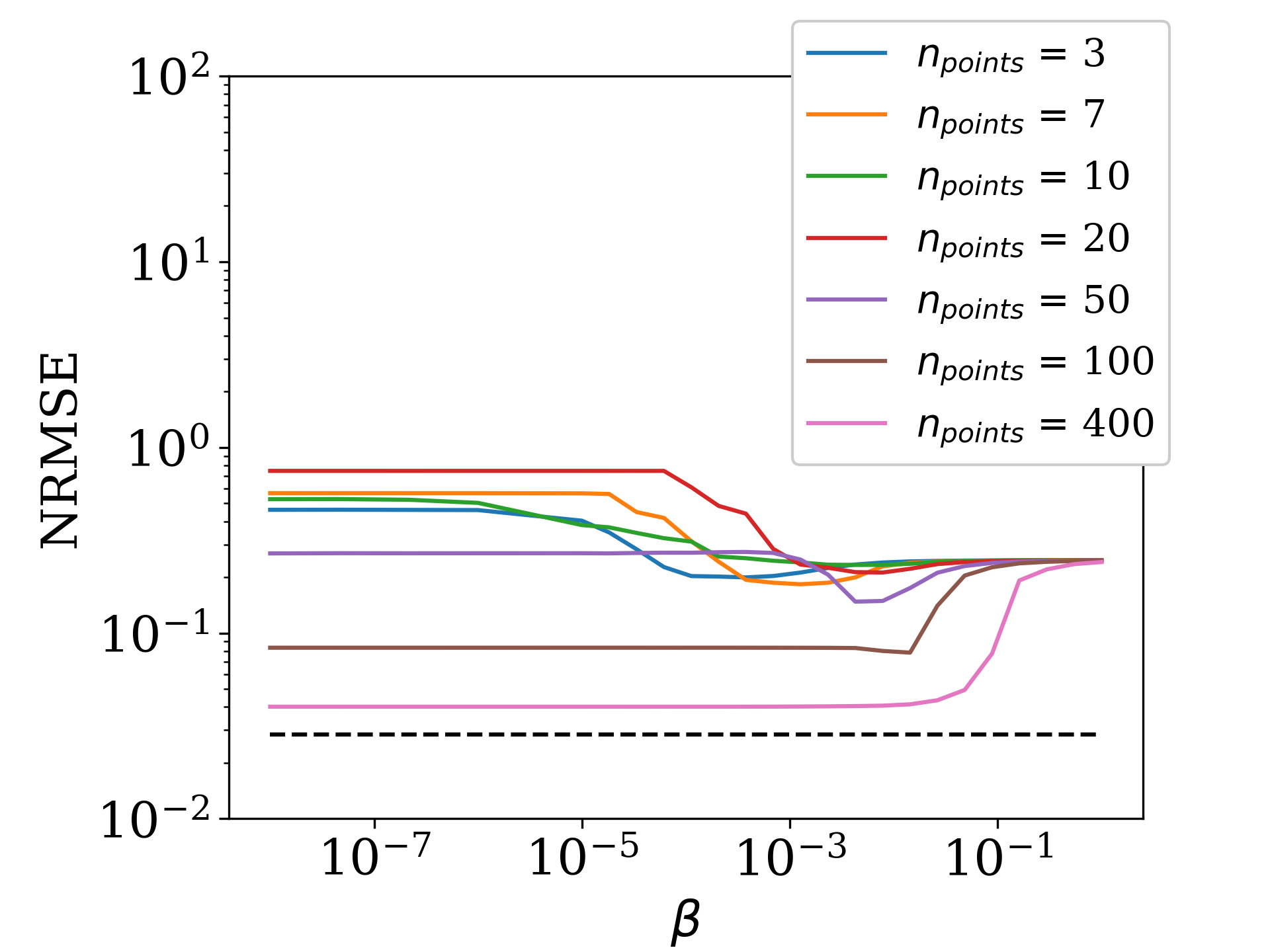}
  \caption{C1 fuel}
\end{subfigure}
\vfill
\begin{subfigure}[b]{0.475\textwidth}
  \centering
  \includegraphics[width=\textwidth]{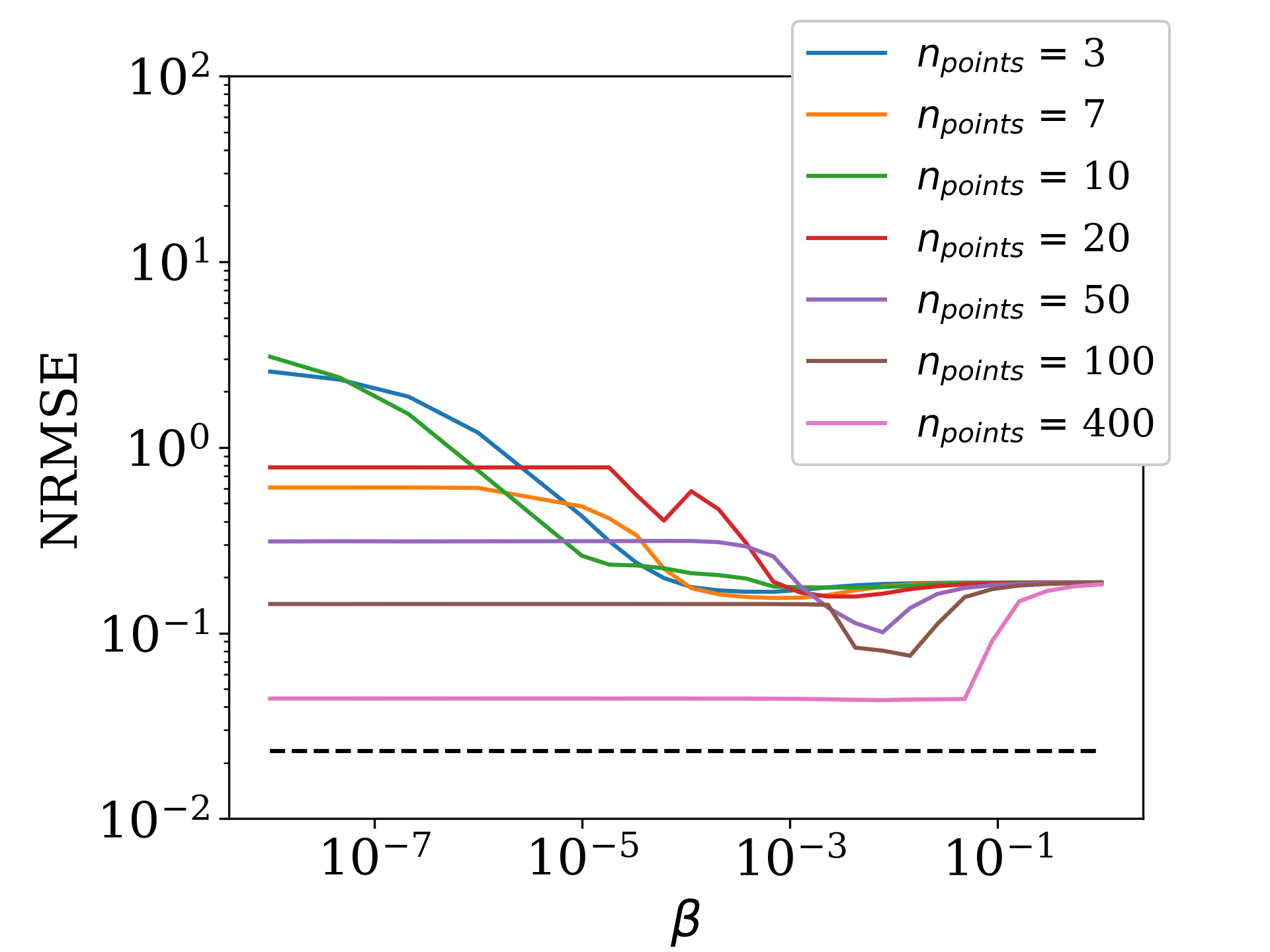}
  \caption{TPRF fuel}
\end{subfigure}
\caption{NRMSE for the data reconstruction of the entire data set using different number of points in the target task.}
\label{fig:l2norm}
\end{figure}

The H$_2$O$_2$ reconstruction profile shown in Fig.~\ref{fig:1D_7points} indicates an optimal amount of knowledge transferred ($\beta$ value) to provide the lowest reconstruction error. The optimal value is between the traditional transfer learning approaches: no transfer or full transfer. As pointed out previously, negative transfer learning can occur when too much knowledge is transferred from a source task that does not fully represent the features in the target task. On the other hand, predictions without transfer learning suffer from the lack of sufficient data to correctly represent the lower dimensional manifold. In order to investigate the effects of knowledge transferred between tasks, Fig.~\ref{fig:l2norm} presents the reconstruction error in terms of the NRMSE for a different number of snapshots used in the target task training for all three fuels employed in this study. The NRMSE was computed for the entire thermochemical vector using all the data points in the original flame solution obtained with Cantera. The NRMSE was computed for the normalized data so that differences in the order of magnitude between different species and temperatures are removed. The dashed horizontal line in Fig.~\ref{fig:l2norm} corresponds to the analytical error when 99.9 \% of the variance is captured for abundant data. 

The reconstruction error increases as the number of snapshots in the target task is reduced for the no transfer learning setting, $\beta \to 0$, as expected. The data sparsity can be quantified by the deviation from the analytical error as the number of training points is reduced for $\beta \to 0$. Tests for $\beta \to 0$, not shown here, reveal that the TPRF predictions approach the analytical error as $n_{points}$ is increased. A significant drop in the NRMSE is observed for $\beta$ values in the order of $10^{-4}$ to $10^{-2}$ depending on the number of training points provided to the target task and the complexity in the chemical mechanism. As $\beta$ approaches unity, all cases with a different amount of data sparsity tend to have the same reconstruction error related to the enforcement of source task knowledge. Negative transfer learning was observed when using a large $n_{points}$ with $\beta \to 1.0$.

\begin{figure}[!ht]
\centering
\begin{subfigure}[b]{0.475\textwidth}
  \centering
  \includegraphics[width=\textwidth]{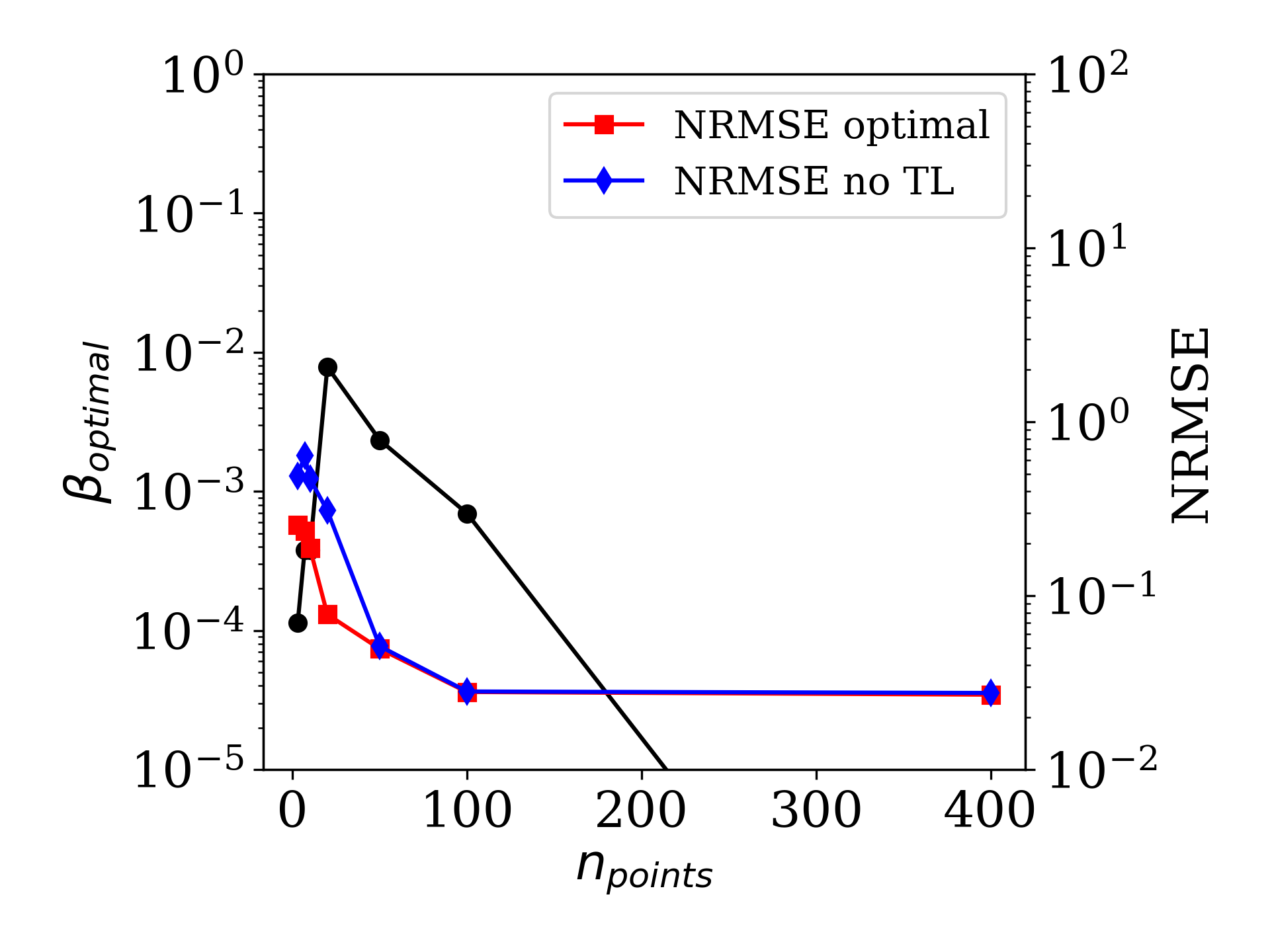}
  \caption{Ammonia-H$_2$ fuel}
\end{subfigure}
\hfill
\begin{subfigure}[b]{0.475\textwidth}
  \centering
  \includegraphics[width=\textwidth, keepaspectratio]{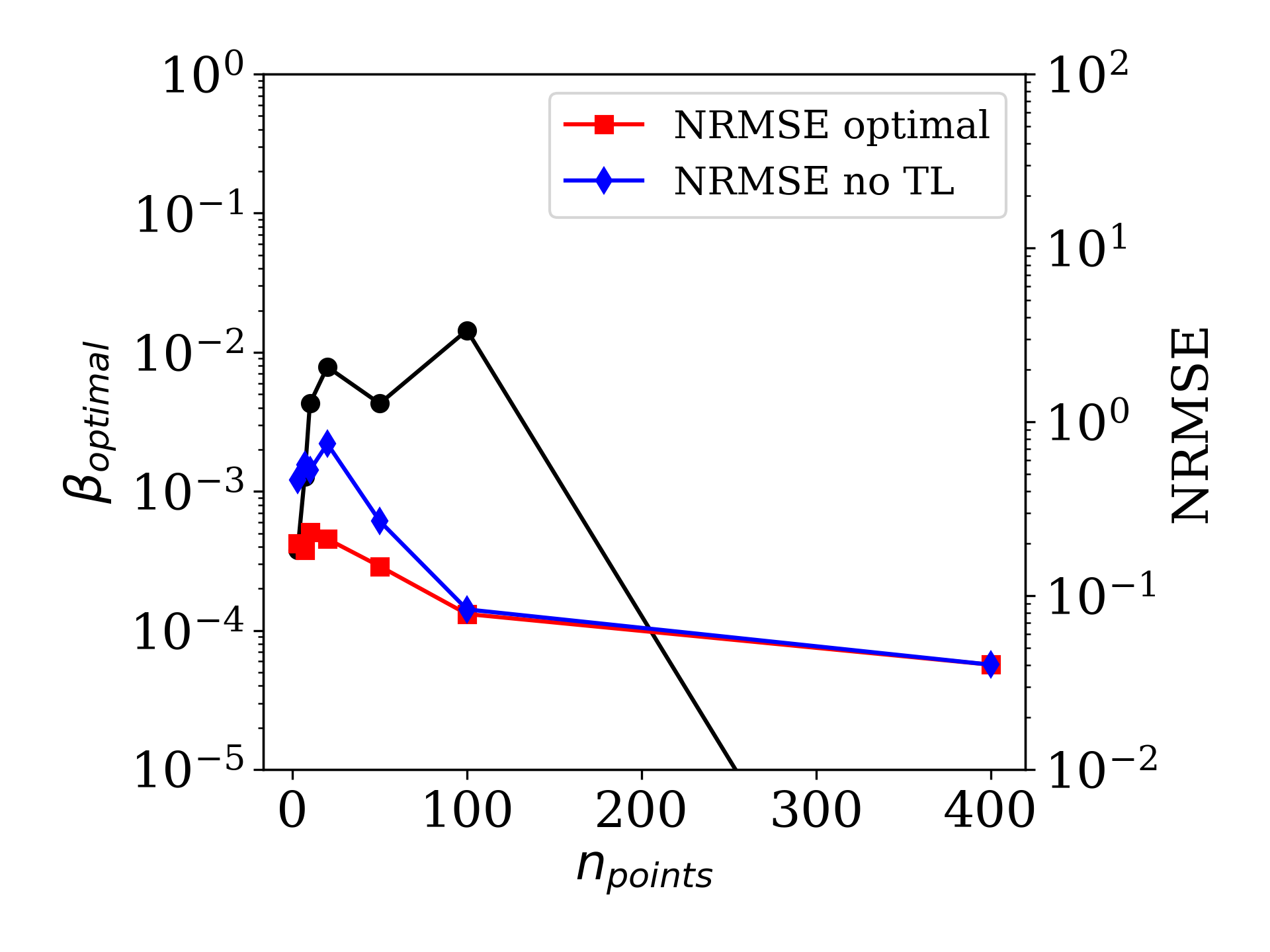}
  \caption{C1 fuel}
\end{subfigure}
\vfill
\begin{subfigure}[b]{0.475\textwidth}
  \centering
  \includegraphics[width=\textwidth]{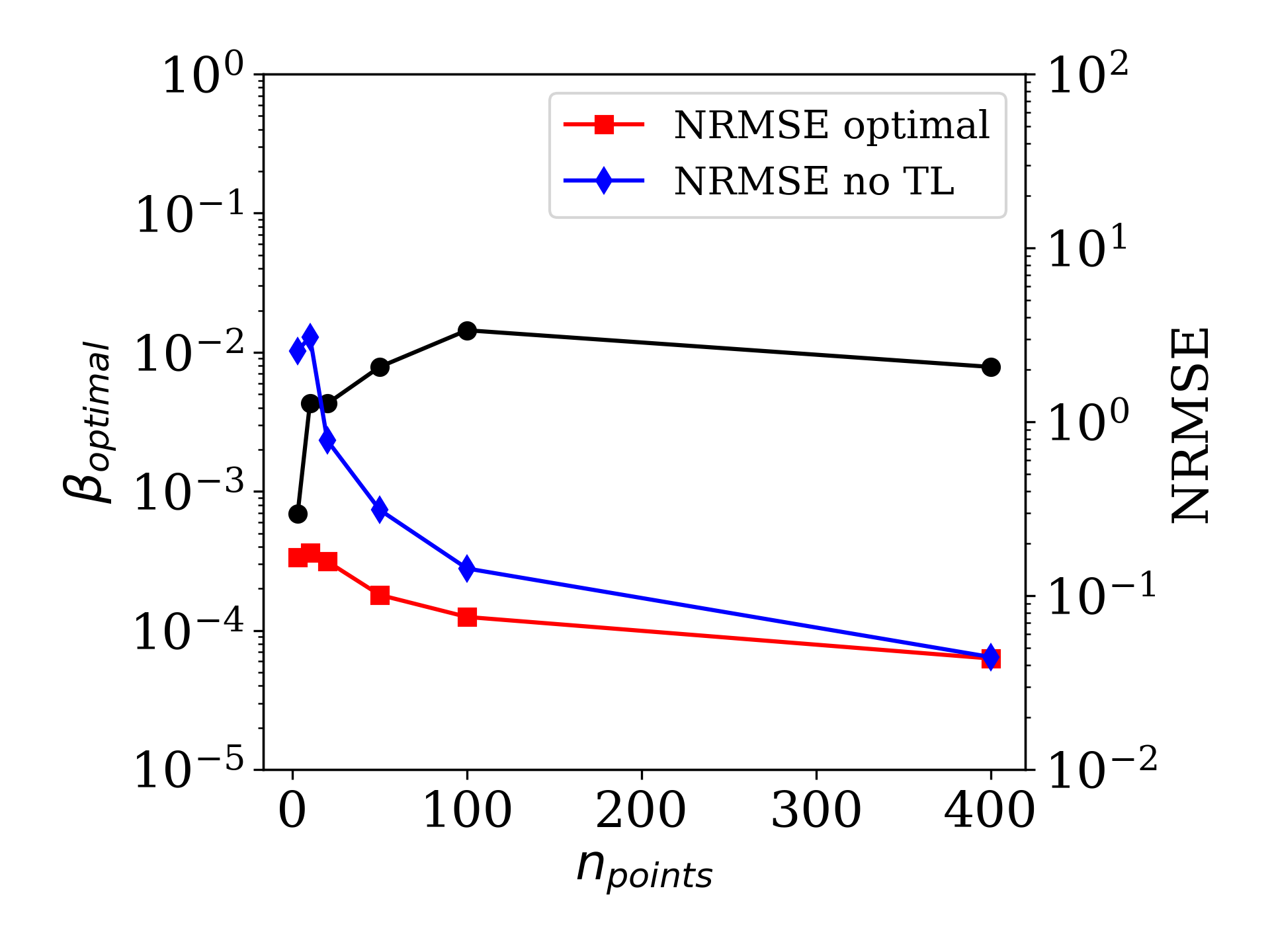}
  \caption{TPRF fuel}
\end{subfigure}

\caption{Variation of optimal transfer learning parameter ($\beta_{optimal}$) as a function of number snapshots used for training (black line).}
\label{fig:beta_opt}
\end{figure}

Figure \ref{fig:beta_opt} presents the optimal transfer learning $\beta_{optimal}$ for a given number of data points in the target task and the corresponding NRMSE. The NRMSE with no transfer learning corresponds to the BNN solution when $\beta \to 0$. The results for the ammonia-H$_2$ and C1 chemical mechanisms present a similar trend in the $\beta_{optimal}$ variation with $n_{points}$. The maximum transfer learning is achieved for the 7-20 points range followed by a significant drop as the number of points is reduced even further. Despite the small magnitude of $\beta_{optimal}$ for $n_{points} \to 0$, the NRMSE is reduced by a factor of 2 when compared to the case with no transfer learning. The TPRF flames present a lower optimal NRMSE throughout the $n_{points}$ space and one order of magnitude reduction in NRMSE is achieved for $n_{points} = 3$ when compared with no transfer learning. It should be noted that the number of points used for training is lower than the number of retained PCs, hence PCA cannot be applied for such data sparsity. Nonetheless, the autoencoders with partial transfer learning present a good performance in capturing the flame profiles.

\subsection{Transfer learning for sparsity in equivalence ratio space}\label{equi_sparse}

\begin{figure}[h!]
    \centering
    \includegraphics[width=3.5in,keepaspectratio=true]{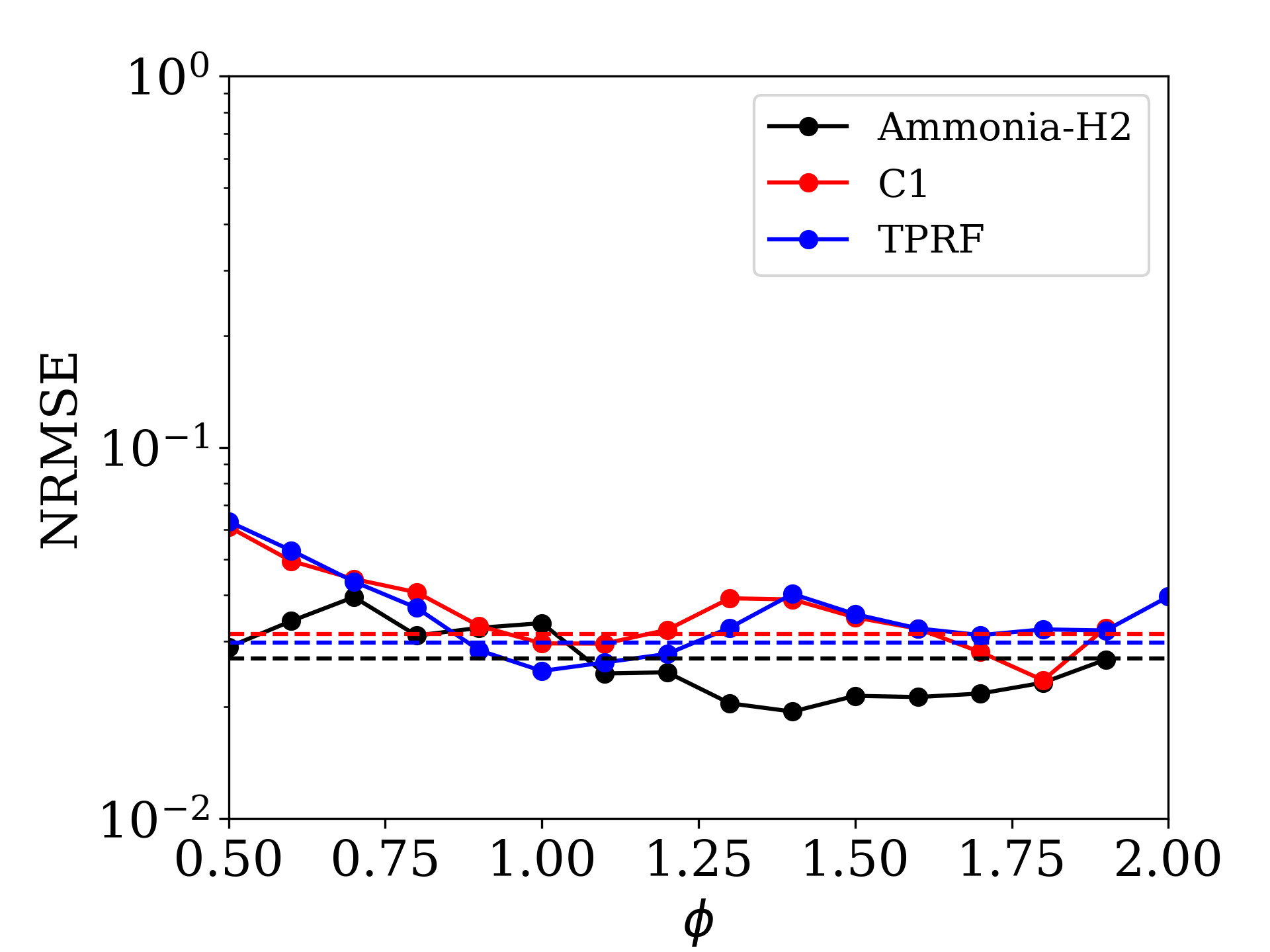}
    \caption{Reconstruction error (NRMSE) as a function of flame equivalence ratio for the source task ($T_u$ = 400 K). Dashed lines correspond to the NRMSE error given the amount of variance captured.}
    \label{fig:l2norm_vs_phi}
\end{figure}

Transfer learning is also applied in the situation where a parametric study is performed, i.e. a data set is obtained for a given set of conditions and a multi-dimensional PC transport simulation is desired at different but similar conditions. The source task for the analysis presented here corresponds to 15 freely-propagating flames with an unburned temperature of 400 K and equivalence ratios ranging from 0.5 to 2.0. The target task corresponds to flames at an unburned temperature of 300 K and data sparsity is created by using a reduced number of equivalence ratios. Figure \ref{fig:l2norm_vs_phi} presents the distribution of the reconstruction error in terms of the NRMSE along the equivalence ratio space for the source task. The NRMSE was computed using Eq.~(\ref{nrmse}) for the reconstruction of each individual equivalence ratio solution using the PC modes from PCA analysis of the entire source task data set. The NRMSE remains stable throughout the equivalence ratio space, therefore the reconstruction error of the target task with equivalence ratio sparsity is not a function of the equivalence ratio selected for the optimization. Small differences in NRMSE among the chemical mechanisms are related to the selection of the subset of PC modes, nonetheless, in all three fuels PCA is capturing at least 99.9 \% of the variance in the data.

\begin{figure}[H]
\centering
\begin{subfigure}[b]{0.475\textwidth}
  \centering
  \includegraphics[width=\textwidth]{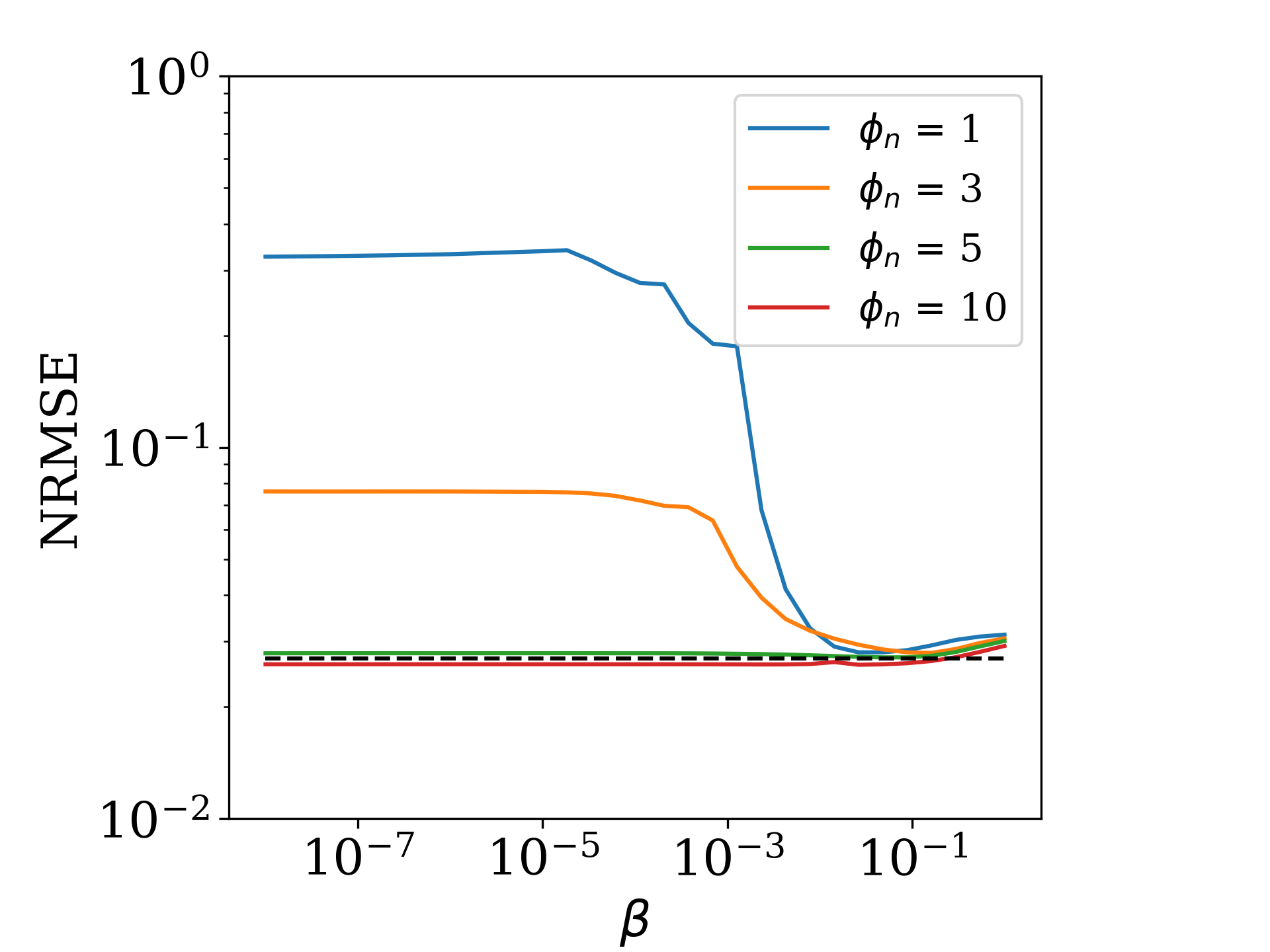}
  \caption{Ammonia-H$_2$ fuel}
\end{subfigure}
\hfill
\begin{subfigure}[b]{0.475\textwidth}
  \centering
  \includegraphics[width=\textwidth, keepaspectratio]{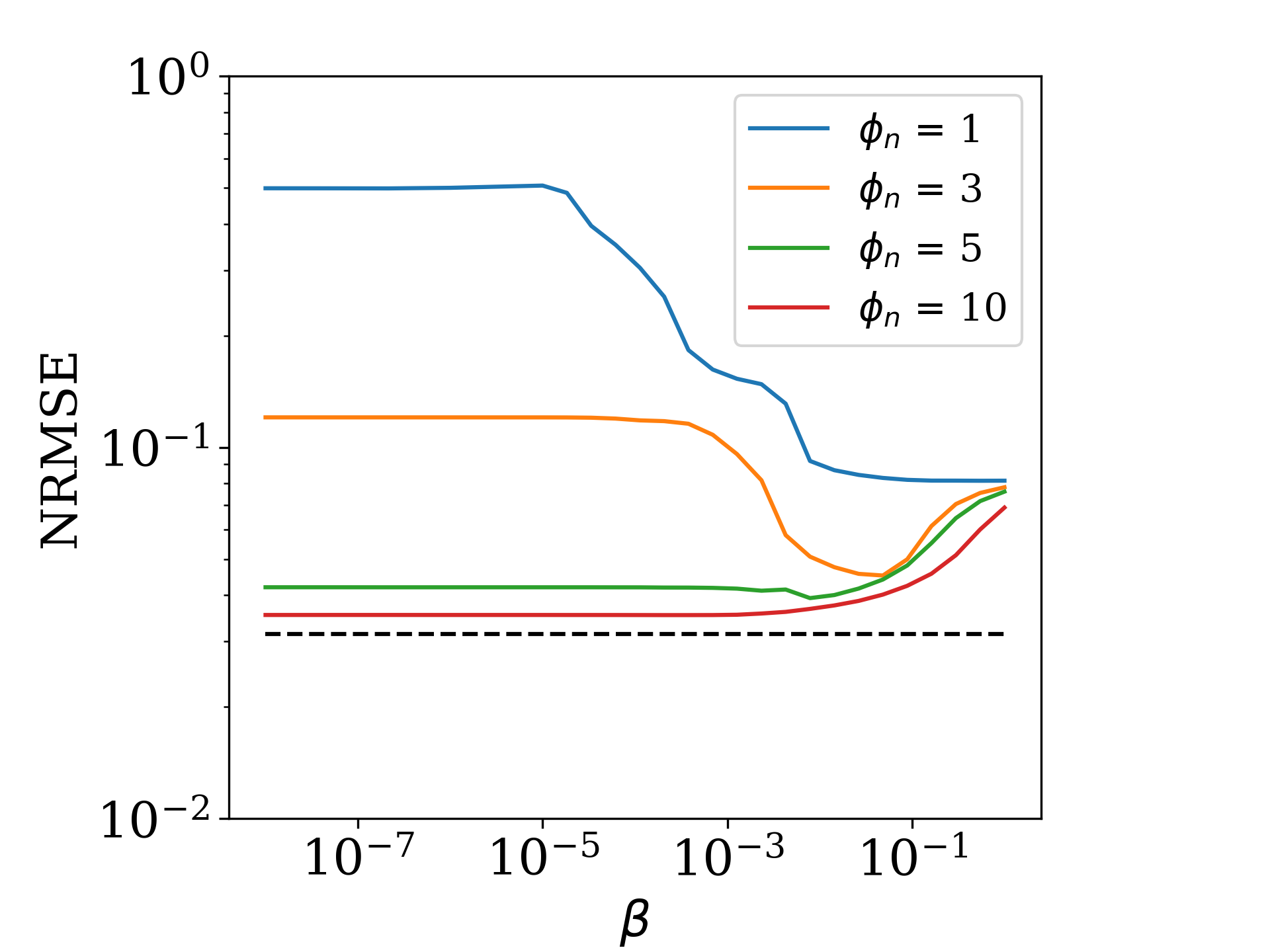}
  \caption{C1 fuel}
\end{subfigure}
\vfill
\begin{subfigure}[b]{0.475\textwidth}
  \centering
  \includegraphics[width=\textwidth]{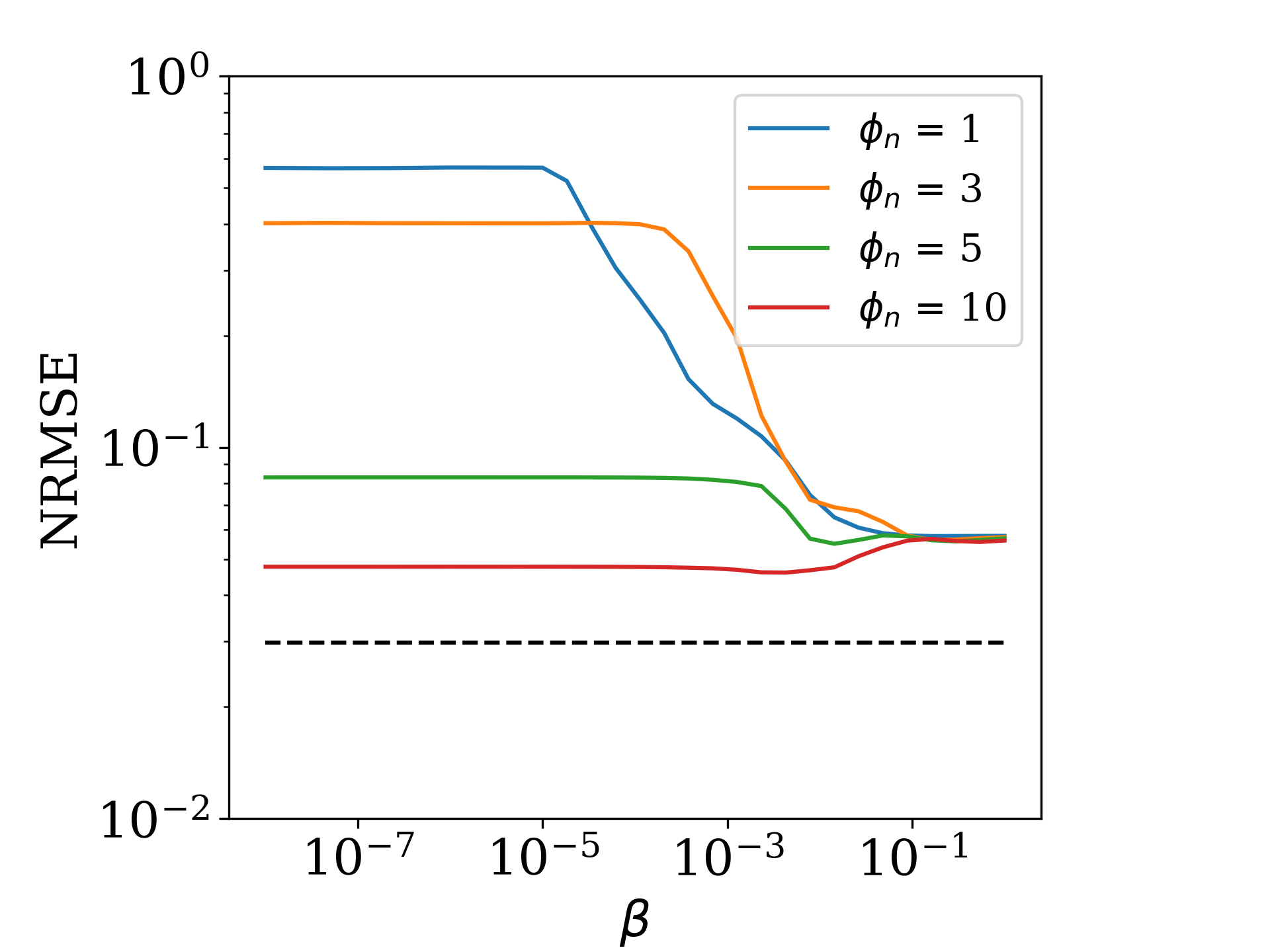}
  \caption{TPRF fuel}
\end{subfigure}
\caption{Reconstruction error (NRMSE) as a function of the transfer learning parameter $\beta$ for different number of equivalence ratios ($\phi_n$) in the target task. NRMSE is computed using target task conditions with $\phi_n = 15$ (all data set). Dashed line corresponds to the NRMSE error given the amount variance captured.}
\label{fig:l2norm_phi}
\end{figure}

Figure \ref{fig:l2norm_phi} shows the reconstruction error in terms of the NRMSE as a function of the transfer learning parameter $\beta$. A different number of premixed flames with different equivalence ratios ($\phi_n$) were used to train the autoenconders for the target task with $T_u$ = 300 K. Similarly to the results presented in Fig.~\ref{fig:l2norm}, the NRMSE was computed by reconstructing the thermochemical vector using all target task data ($\phi_n = 15$). The dashed line in Fig.~\ref{fig:l2norm_phi} corresponds to the reference modeling error in the dimensionality reduction of the problem considering the amount of variance captured. The results with $\phi_n = 10$ reveal that the autoencoders without transfer learning ($\beta \to 0.0$) can reach the modeling error for ammonia-H$_2$ and C1, as expected. The reconstruction error for the TPRF chemical mechanism does not reach the modeling error for the same parameters, i.e. $\phi_n = 10$ and $\beta \to 0.0$. Three mechanisms are identified to affect the model performance: the amount of data in the source task; the amount of data in the target task; and data similarity. The amount of data in the source task and the data similarity are not the source of uncertainty when there is no transfer learning. The lack of performance can be attributed to the small number of data points to train the autoencoders for a complex manifold as present in the TPRF mechanism. It is worth mentioning that the number of points required is related to the complexity of the manifold and not to the number of species in the mechanism. Tests, not shown here, were performed by reducing the number of dimensions in the training from 33 for the TPRF fuel to 20 and 10 while keeping the same number of data samples. The test effectively reduces the number of data points required to correctly capture the lower dimensional manifold. The results showed that the reconstruction error starts to approximate the minimum modeling error as the number of dimensions is reduced.

The reduction in data reconstruction error with the number of equivalence ratios used for training is a function of the similarity in manifold space between different equivalence ratios for each fuel. The more complex fuels are observed to have a larger NRMSE as the number of $\phi_n$ is reduced for training with $\beta \to 0.0$. Nonetheless, the reduction in $\phi_n$ for the target task yields an increase in the reconstruction error for all fuels, as expected. The predictions for $\phi_n = 10$ and $\beta \to 0.0$ approach the modeling optimal performance and therefore no reduction in data reconstruction error is observed when TL is applied. In fact, an increase in NRMSE is observed when $\beta \to 1.0$ denoting a negative transfer knowledge. Transfer learning starts to play a role in the reduction of the NRMSE when data sparsity is increased. The large data sparsity case with $\phi_n = 1$ shows an order of magnitude reduction in the NRMSE when $\beta > 10^{-3}$ for all chemical mechanisms tested. The solution with $\phi_n = 3$ corresponds to an intermediate data sparsity scenario where a $\beta$ value between 0 and 1 approximates the reconstruction error obtained with $\phi_n = 10$. All curves approach the same NRMSE magnitude when $\beta \to 1$ as a consequence of the full transfer of knowledge. Similarly to the results presented in Section \ref{sec:spatial_sparse}, the evolution in NRMSE with $\beta$ shows that there is an optimal amount of transfer learning $\beta_{optimal}$ to be performed in order to obtain the minimum reconstruction error. The $\beta_{optimal}$ is mainly a function of the data sparsity, complexity in the lower dimensional manifold and the similarity between tasks.

\begin{figure}[H]
\centering
\begin{subfigure}[b]{0.475\textwidth}
  \centering
  \includegraphics[width=\textwidth]{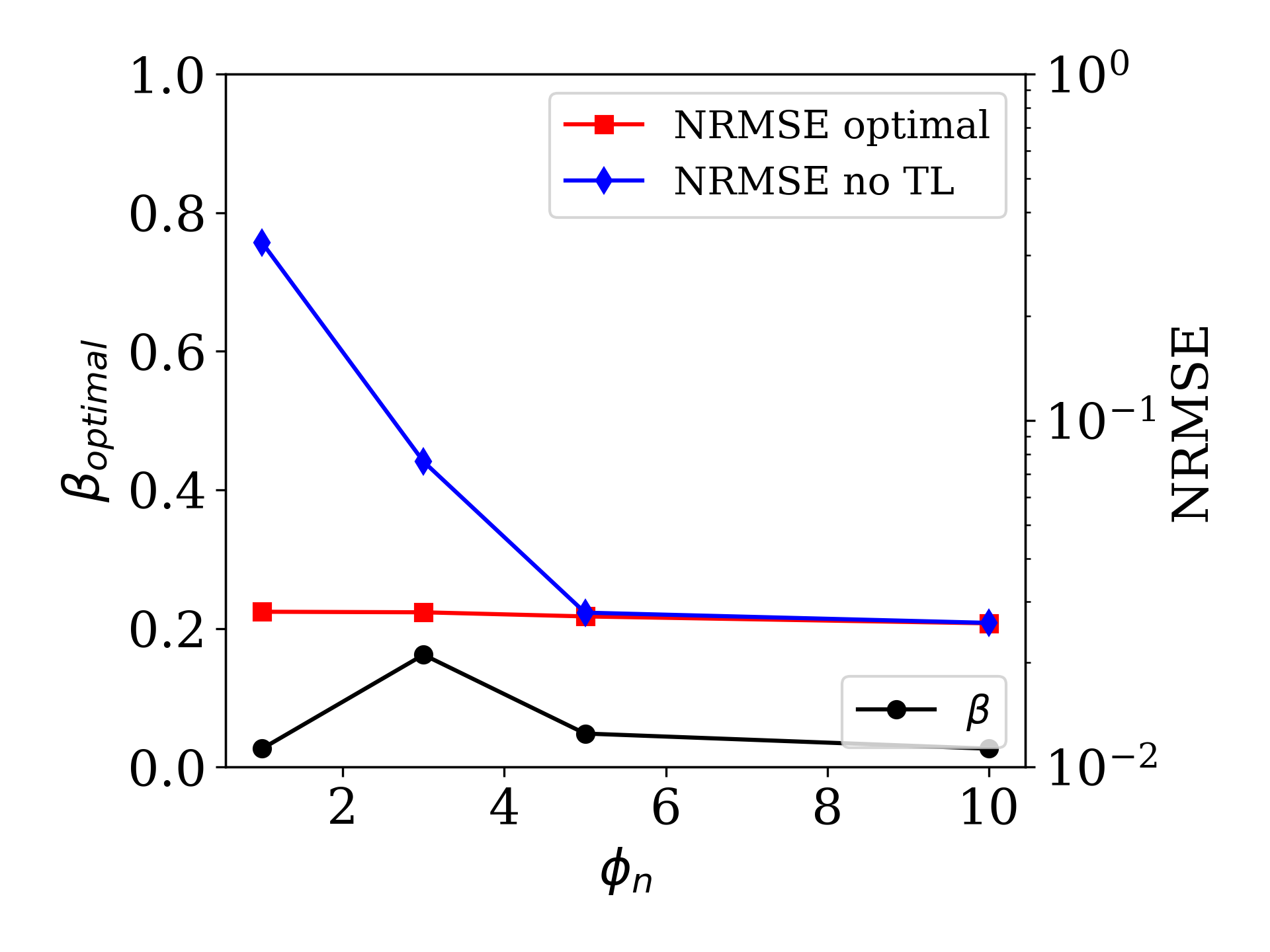}
  \caption{Ammonia-H$_2$ fuel}
\end{subfigure}
\hfill
\begin{subfigure}[b]{0.475\textwidth}
  \centering
  \includegraphics[width=\textwidth, keepaspectratio]{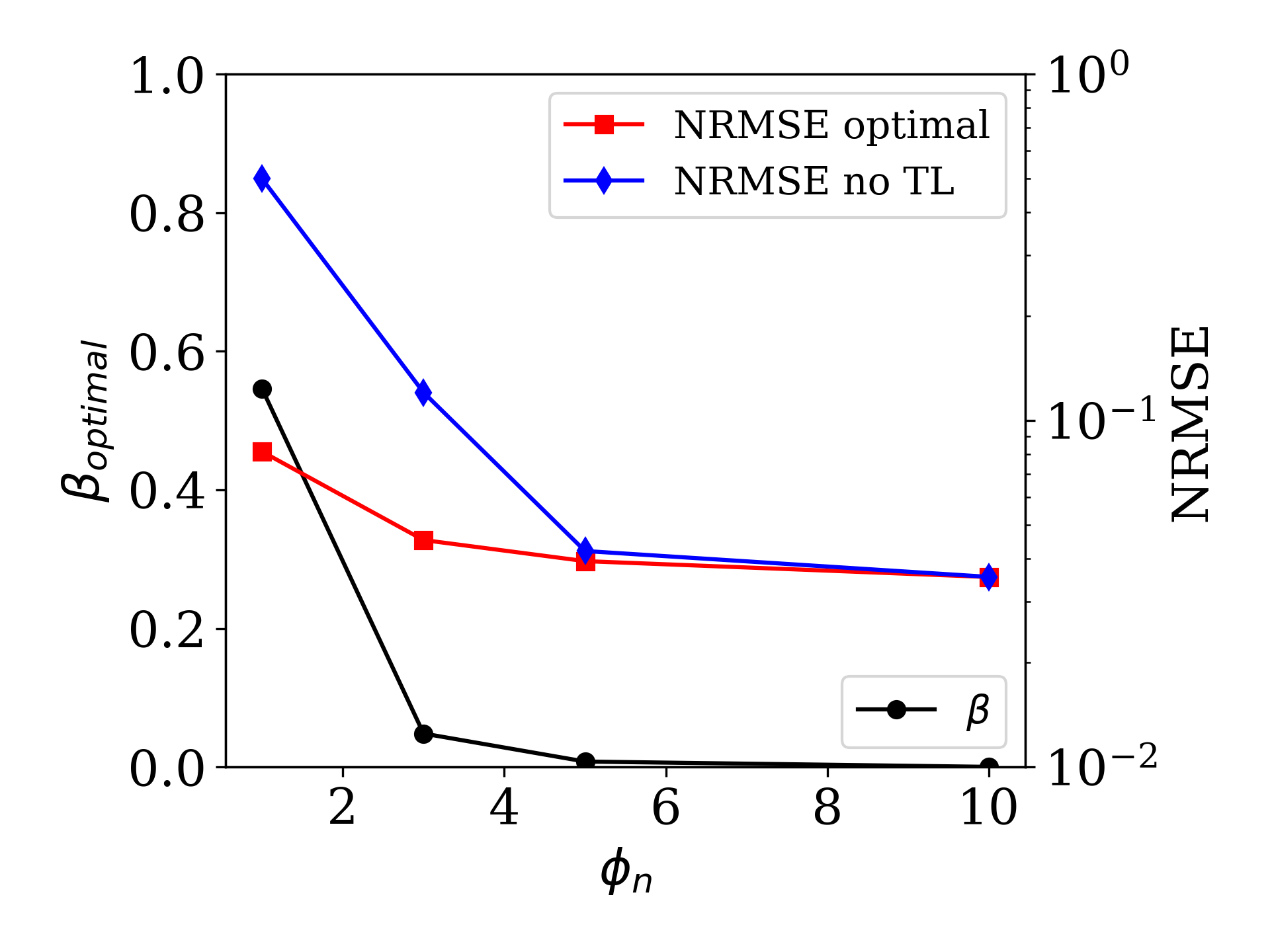}
  \caption{C1 fuel}
\end{subfigure}
\vfill
\begin{subfigure}[b]{0.475\textwidth}
  \centering
  \includegraphics[width=\textwidth]{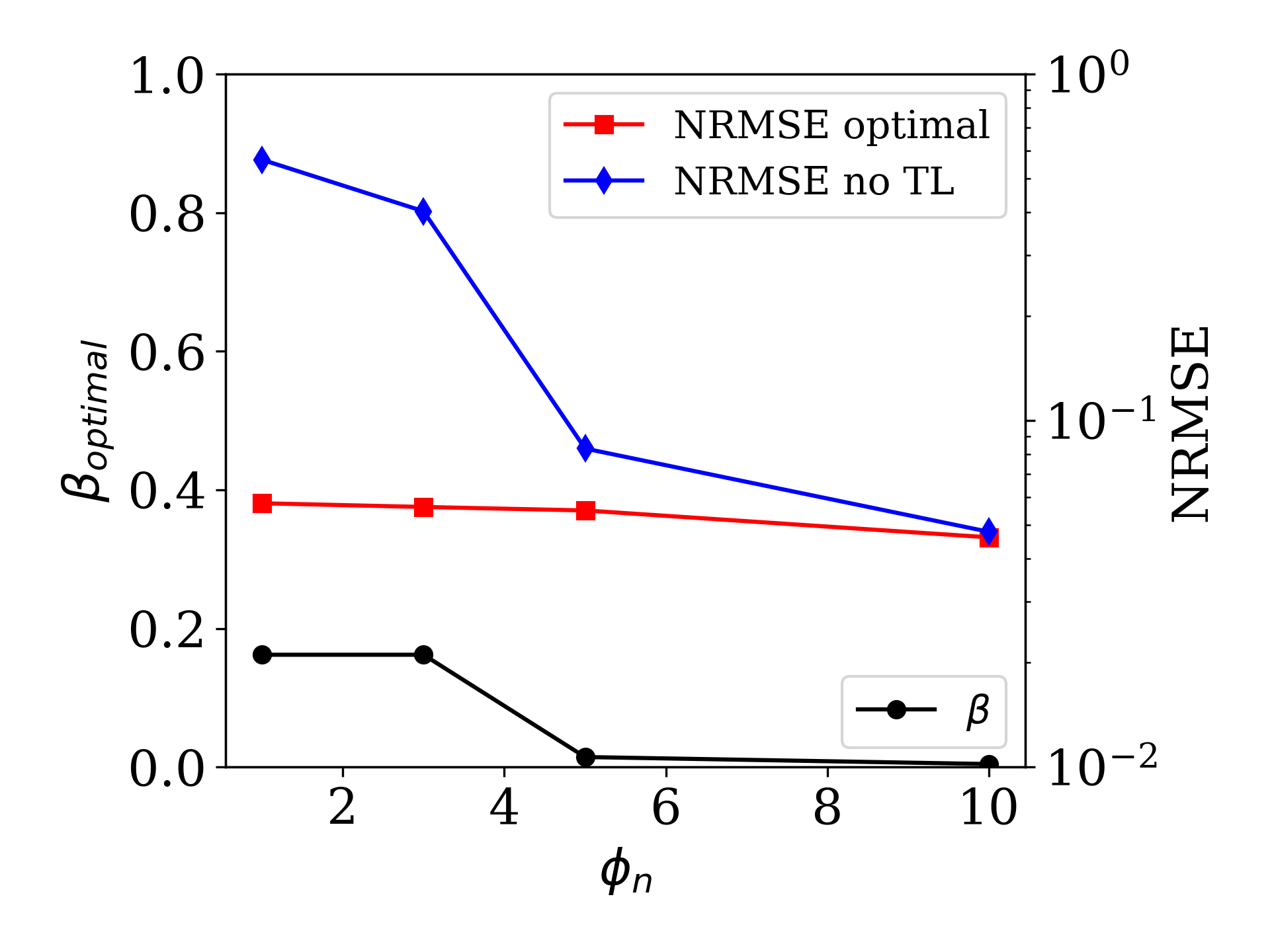}
  \caption{TPRF fuel}
\end{subfigure}
\caption{Optimal $\beta$ and NRMSE as a function of number of equivalence ratios in the target task.}
\label{fig:betaOpt_phi}
\end{figure}

Figure \ref{fig:betaOpt_phi} presents the optimal value for the transfer learning parameter $\beta$ which corresponds to the minimum NRMSE for each $\phi_n$ in Fig.~\ref{fig:l2norm_phi}. The corresponding optimal NRMSE and NRMSE without transfer learning ($\beta \to 0$) are also presented. The variation of $\beta_{optimal}$ with respect to data sparsity follows the expected trend in the case where source and target tasks are similar, i.e. the optimal amount of TL tends to increase with data sparsity and no transfer learning is performed in the case of abundant data. The NRMSE increases exponentially with the reduction in $\phi_n$ for the case with no transfer learning ($\beta = 0.0$), whereas with partial transfer learning it remains steady across different $\phi_n$'s. The small variation in the NRMSE with the reduction in $\phi_n$ highlights the performance of the transfer learning strategy developed in this study. In order to obtain the same NRMSE as for $\beta_{optimal}$ with $\phi_n = 1$, the ammonia-H$_2$ and C1 fuels with no transfer learning would require approximately 4-5 times more data. The benefits of partial transfer learning are more pronounced for a more complex manifold, with the TPRF requiring approximately 10 times more data to reproduce the reconstruction error of $\phi_n = 1$.

\begin{figure}[H]
\centering
\begin{subfigure}[b]{0.475\textwidth}
  \centering
  \includegraphics[width=\textwidth]{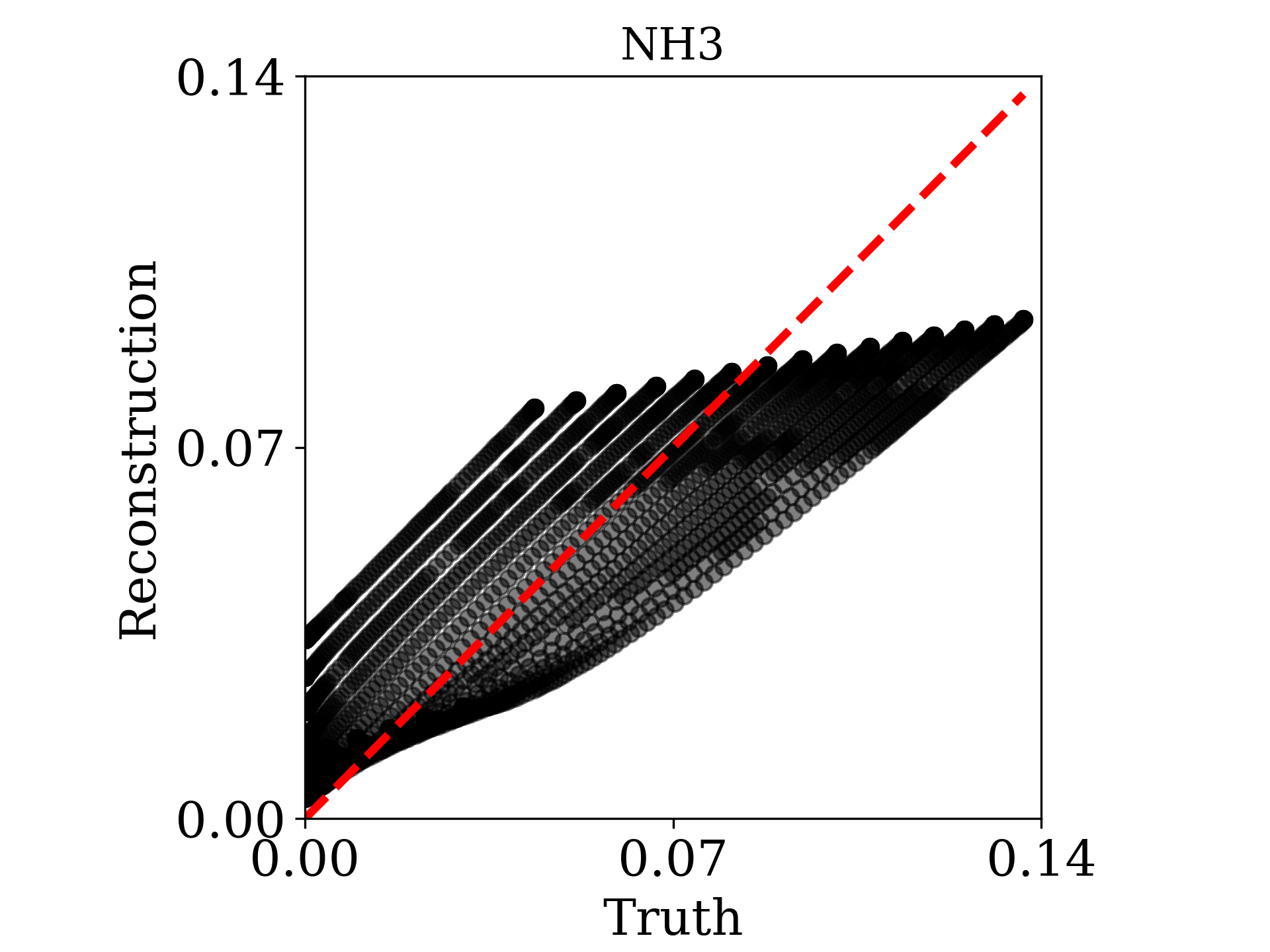}
  \caption{$\beta = 0.0$}
\end{subfigure}
\hfill
\begin{subfigure}[b]{0.475\textwidth}
  \centering
  \includegraphics[width=\textwidth, keepaspectratio]{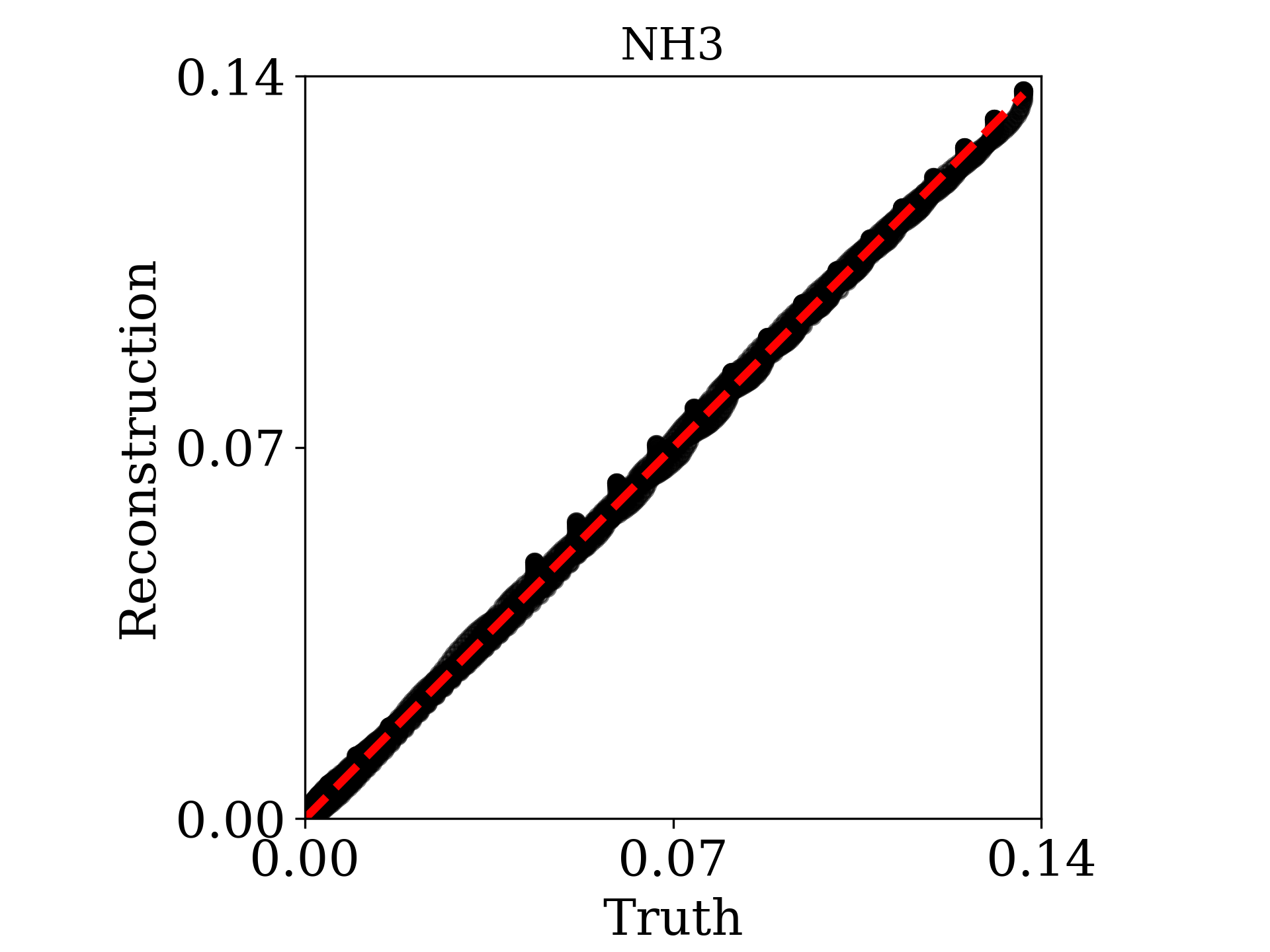}
  \caption{$\beta = \beta_{optimal}$}
\end{subfigure}

\vfill

\begin{subfigure}[b]{0.475\textwidth}
  \centering
  \includegraphics[width=\textwidth]{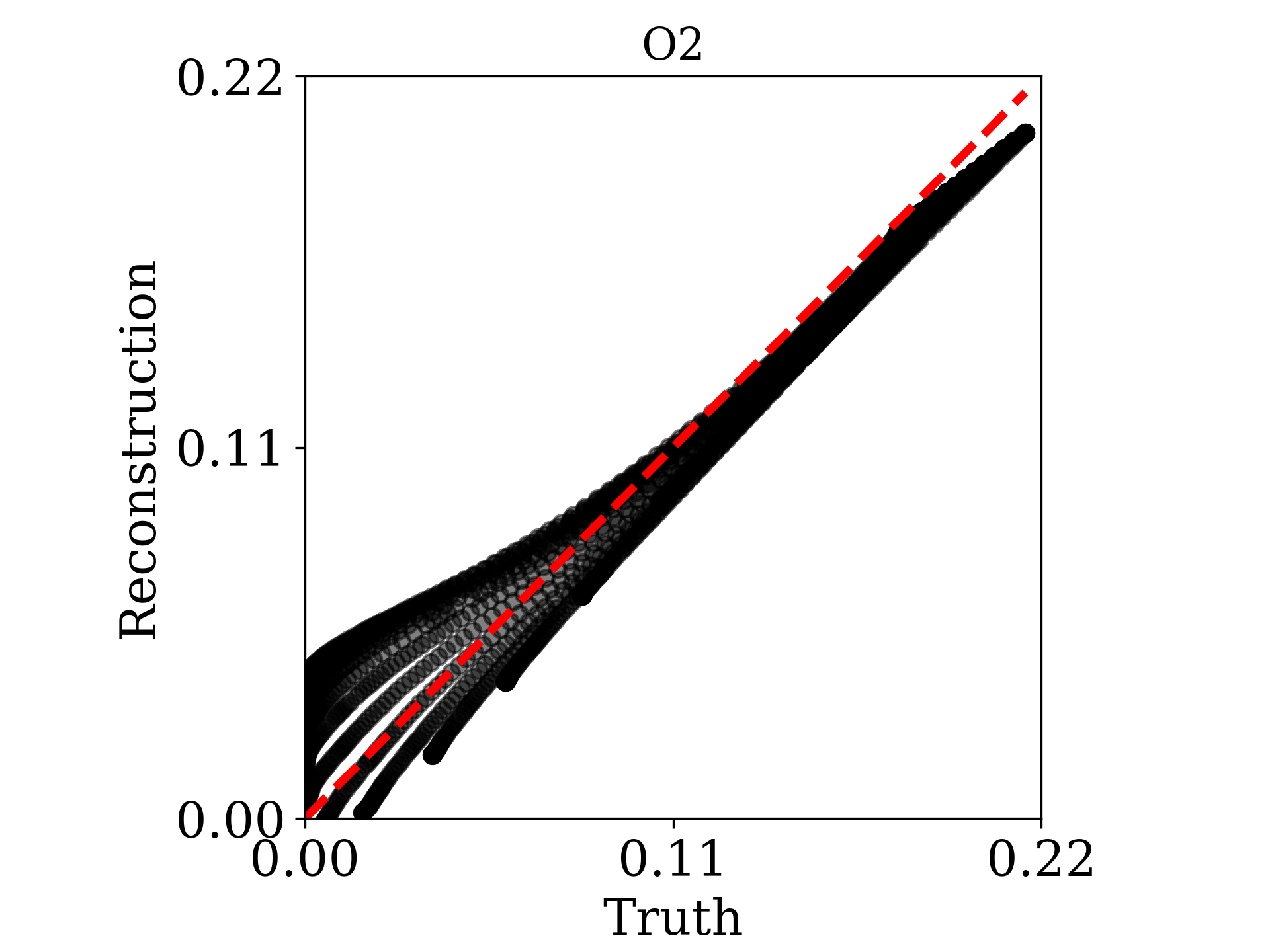}
  \caption{$\beta = 0.0$}
\end{subfigure}
\hfill
\begin{subfigure}[b]{0.475\textwidth}
  \centering
  \includegraphics[width=\textwidth, keepaspectratio]{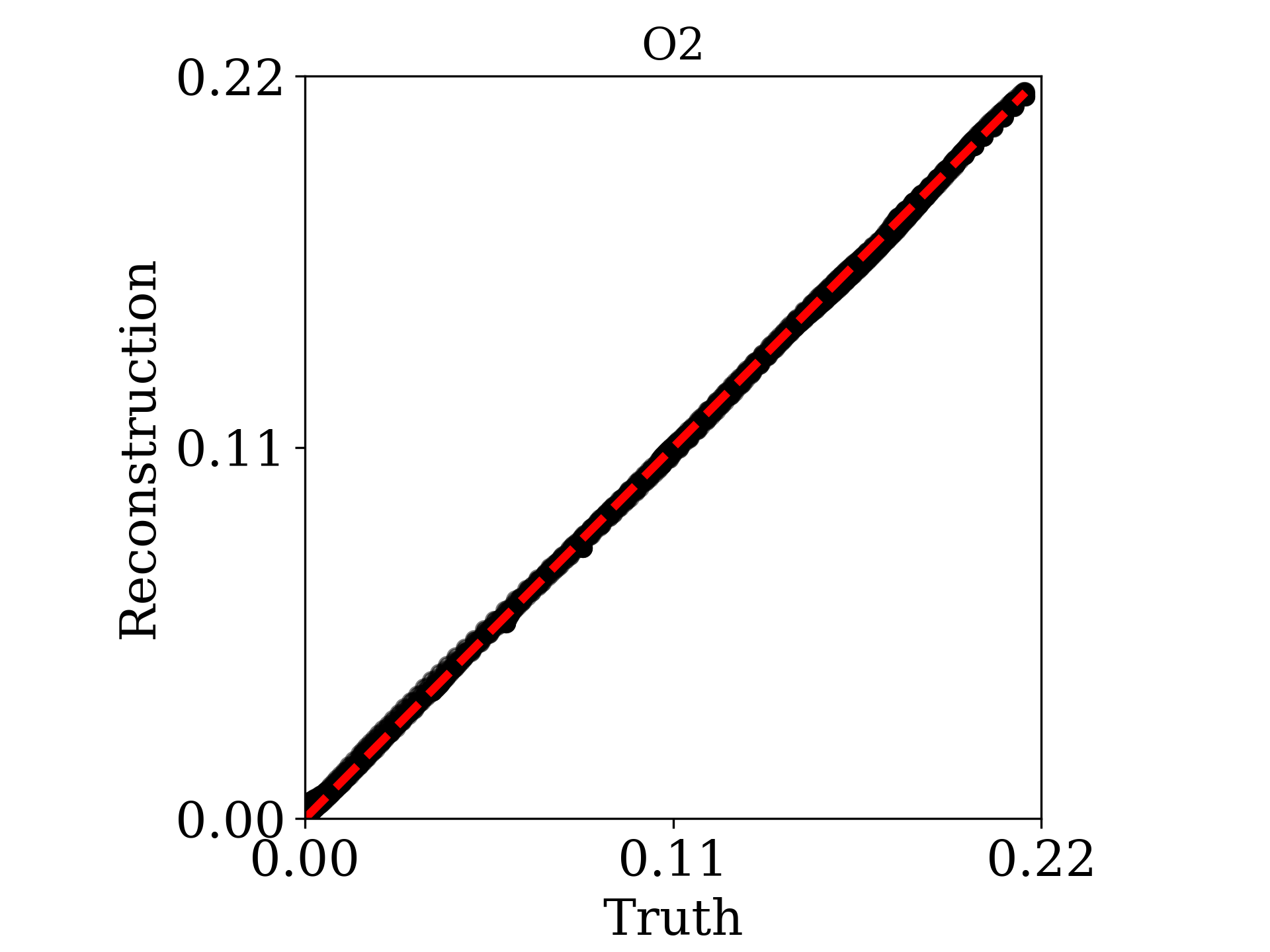}
  \caption{$\beta = \beta_{optimal}$}
\end{subfigure}

\caption{Comparison between data reconstruction and flame solution (truth) for different transfer learning parameters $\beta$ using the 1 equivalence ratio for the target task with the NH$_3$-H$_2$ mechanism. Top row presents NH$_3$ mass fraction; bottom row presents $O_2$ mass fraction. Dashed red line corresponds to the ideal linear behavior.}
\label{fig:reconstruction}
\end{figure}

Figure \ref{fig:reconstruction} shows a comparison in terms of the true and reconstructed fuel and molecular oxygen mass fractions using all equivalence ratios present in the target task ($\phi_n = 15$). The selected scalars correspond to the ammonia-H$_2$ one-dimensional flames at $T_u$ = 300 K (truth) and the reconstruction was performed using the BNN solution for only one equivalence ratio ($\phi_n = 1$) in the target task. The data reconstruction is performed for two different $\beta$, one corresponding to no transfer learning and the other at the optimal $\beta$ presented in Fig.~\ref{fig:betaOpt_phi}a. Figure \ref{fig:reconstruction} shows a clear shift in the reconstructed mass fractions scatter towards the ideal behavior when $\beta_{optimal}$ is used. 

\begin{figure}[H]
\centering
\begin{subfigure}[b]{0.475\textwidth}
  \centering
  \includegraphics[width=\textwidth]{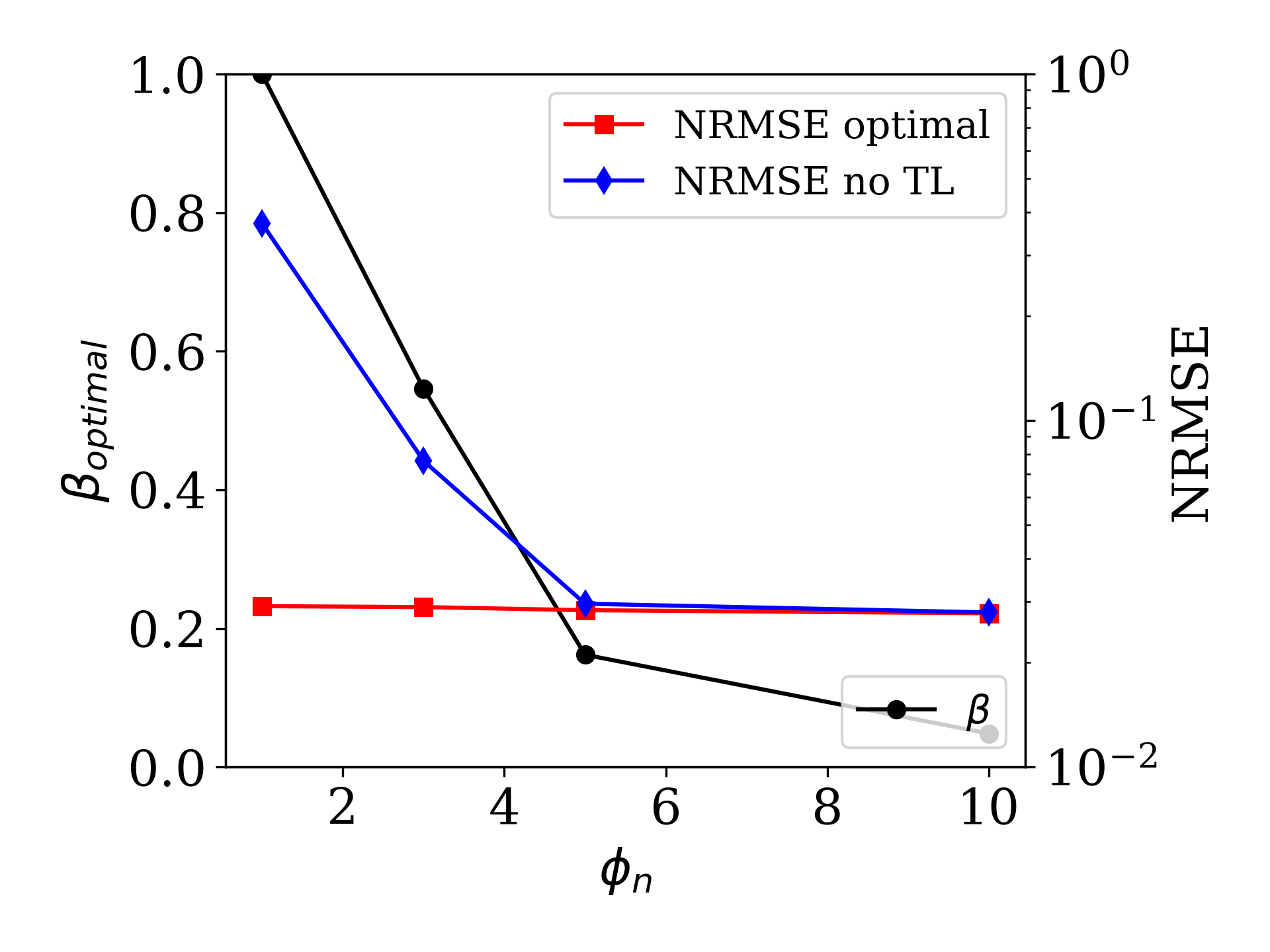}
  \caption{$T_u$ = 450 K}
\end{subfigure}
\hfill
\begin{subfigure}[b]{0.475\textwidth}
  \centering
  \includegraphics[width=\textwidth, keepaspectratio]{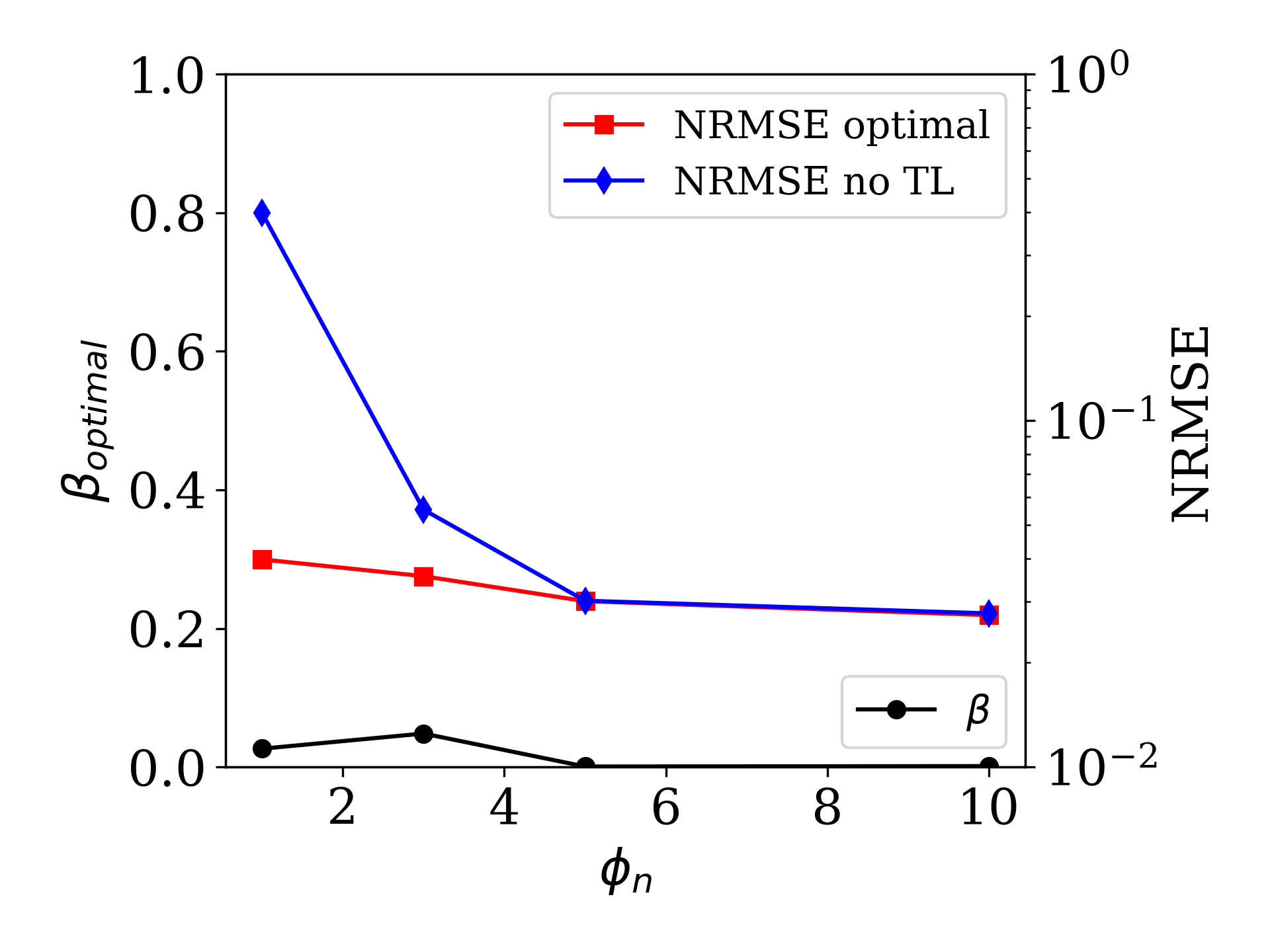}
  \caption{$T_u$ = 600 K}
\end{subfigure}
\caption{Optimal $\beta$ and NRMSE as a function of a number of equivalence ratios for different unburned temperatures for the target task for the NH$_3$-H$_2$ mechanism.}
\label{fig:betaOpt_Tu}
\end{figure}

The results for transfer learning with data sparsity in equivalence ratio space have been presented thus far for oxidizer temperatures of 300 and 400 K, respectively, for target and source tasks. The similarity between source and target tasks and its effects on $\beta_{optimal}$ are analyzed for the ammonia-H$_2$ fuel by changing the unburned temperature of the freely-propagating flame solutions for the target task. An increase in similarity for the target task was obtained by using flame solutions with an oxidizer temperature of 450 K. On the other hand, a reduction in similarity was obtained by setting the target task to have an oxidizer temperature of 600 K. Figure \ref{fig:betaOpt_Tu} shows the NRMSE variation and $\beta_{optimal}$ with respect to the number of equivalence ratios used in the target task. The results show that $\beta_{optimal}$ approaches unity for large sparsity when source and target tasks are similar while maintaining a constant NRMSE. Conversely, the predictions with 600 K show a small value of $\beta_{optimal}$ for large data sparsity while the NRMSE remains steady.

\begin{figure}[H]
\centering
\begin{subfigure}[b]{1.0\textwidth}
  \centering
  \includegraphics[width=\textwidth]{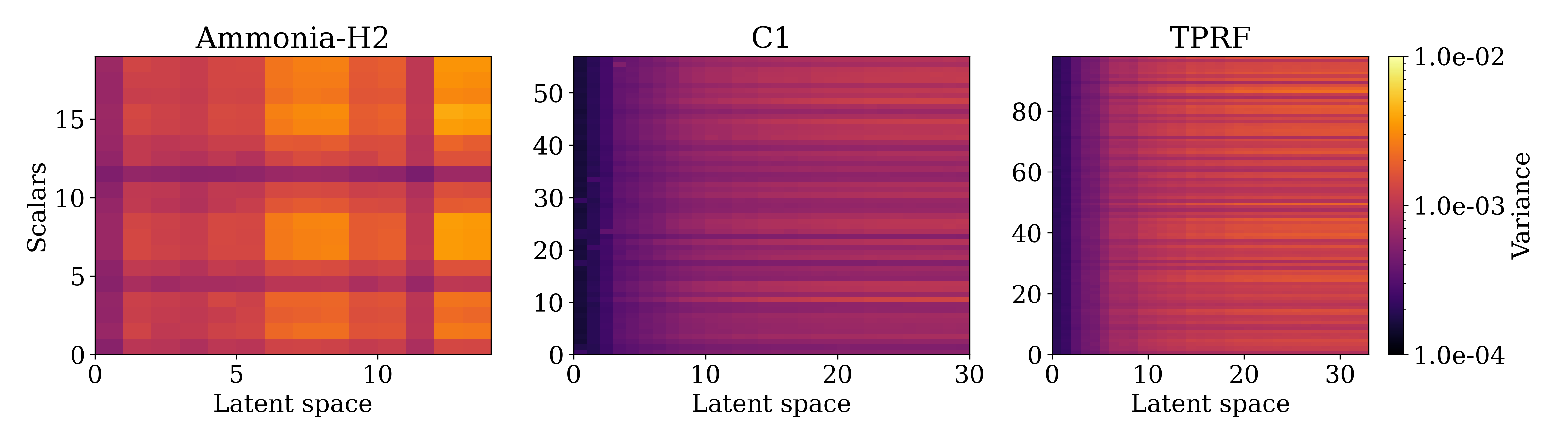}
  \caption{$\phi_n = 1$}
\end{subfigure}
\vfill
\begin{subfigure}[b]{1.0\textwidth}
  \centering
  \includegraphics[width=\textwidth, keepaspectratio]{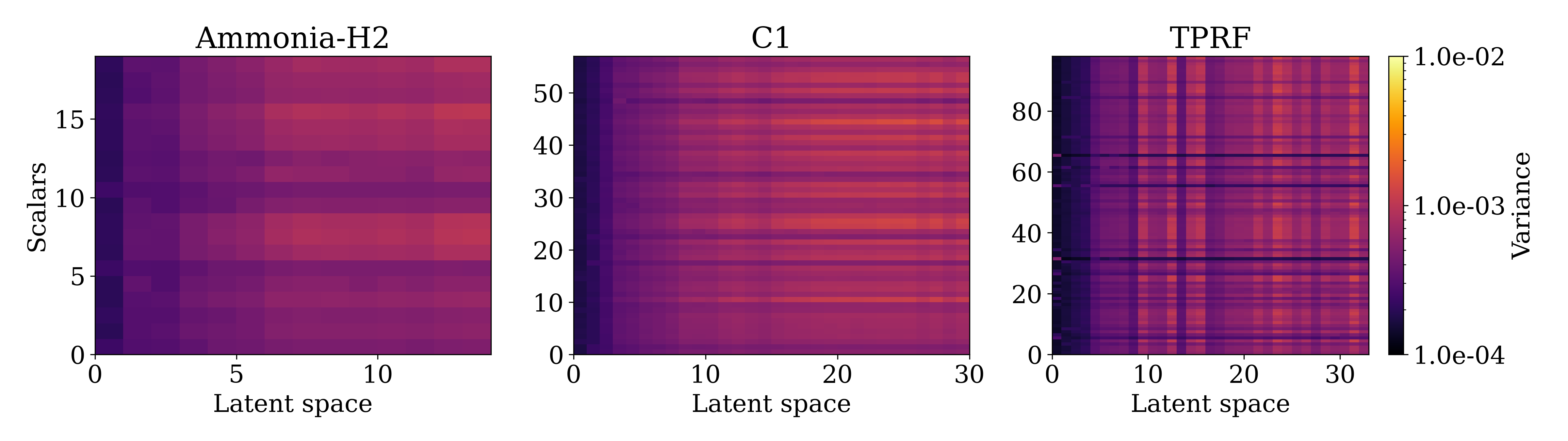}
  \caption{$\phi_n = 10$}
\end{subfigure}
\vfill
\caption{Predicted variance in the autoencoder weights for the target task with $\beta = \beta_{optimal}$}
\label{fig:weights_scale}
\end{figure}

The dimensionality reduction performed by the tied autoencoder with linear activation function is similar to the PCA analysis where the PC modes capture the correlation between scalars. Low-rank PC modes or components capture the highest correlation in the data set whereas higher-rank PC modes capture smaller correlations. The is no constraint for the autoencoder weights (PC modes) to be ordered by the correlation in the data set. However, the initial condition for the optimization using the PC modes obtained from PCA of the target task data gives it an approximate solution ranked by the correlation. Further evidence of ranked modes in the autoencoders will be shown later in the text when presenting Fig.~\ref{fig:weights_scale}. Figure \ref{fig:PC_modes} showed that the first PC mode (Component 1) remains with a similar distribution up to $\phi_n = 3$. As the PC mode rank is increased, the PC values start to deviate from the reference distribution with $\phi_n = 15$. This behavior suggests that the first modes can be captured in a data sparsity setting and do not require transfer learning, only the higher ranks would benefit from the transfer of knowledge. Given the fact that transfer learning is performed in lower dimensional space by modifying the prior distribution, the hypothesis can be tested by simply selecting which priors will be informative. 

\begin{figure}[H]
\centering
\begin{subfigure}[b]{0.475\textwidth}
  \centering
  \includegraphics[width=\textwidth]{figures/h2_nh3/phi_sparsity/beta_optimal_phi_sparsity_300.0.png}
  \caption{Transfer of all PC modes}
\end{subfigure}
\hfill
\begin{subfigure}[b]{0.475\textwidth}
  \centering
  \includegraphics[width=\textwidth, keepaspectratio]{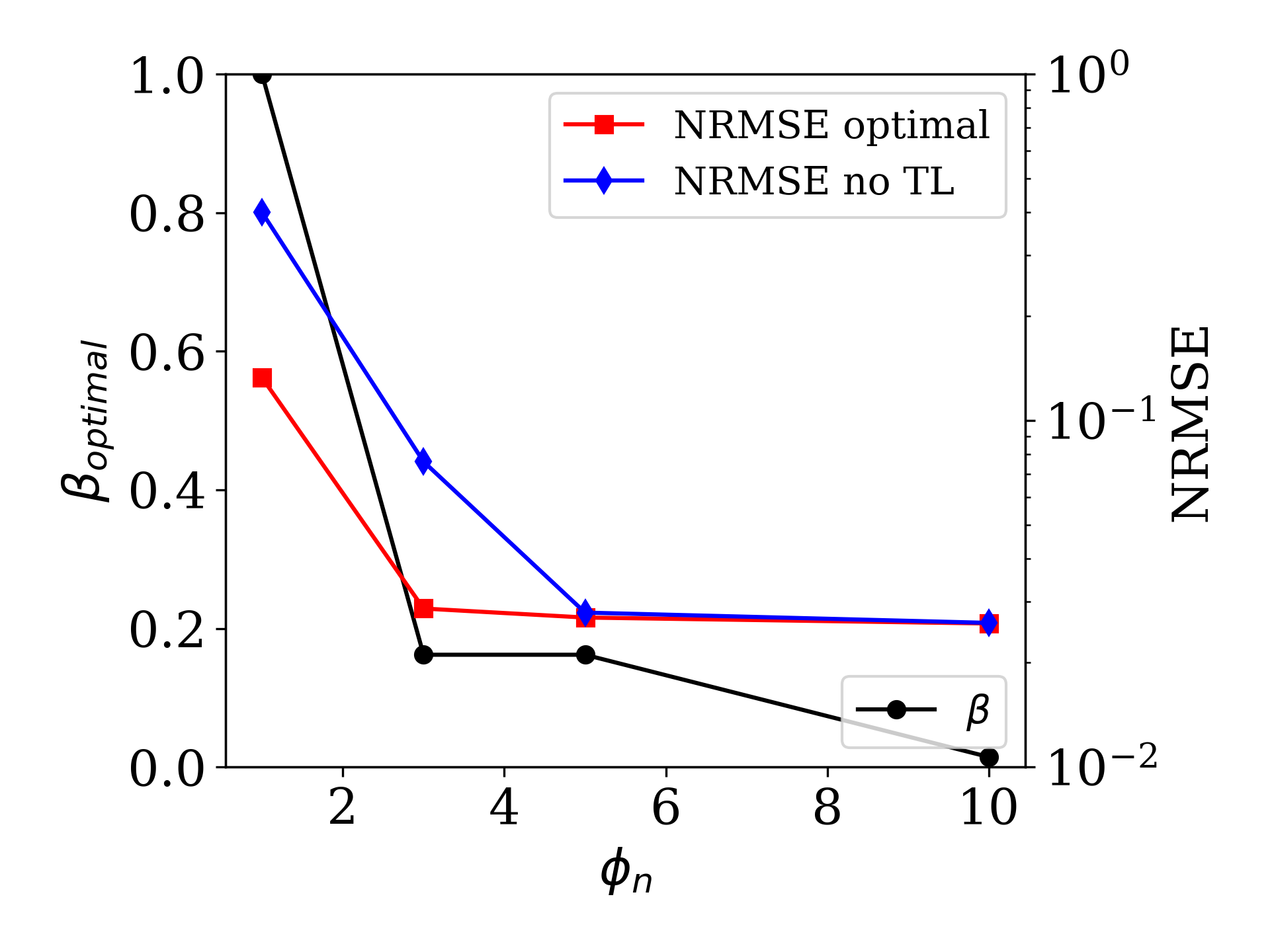}
  \caption{Transfer of the last 10 PC modes}
\end{subfigure}
\vfill
\begin{subfigure}[b]{0.475\textwidth}
  \centering
  \includegraphics[width=\textwidth, keepaspectratio]{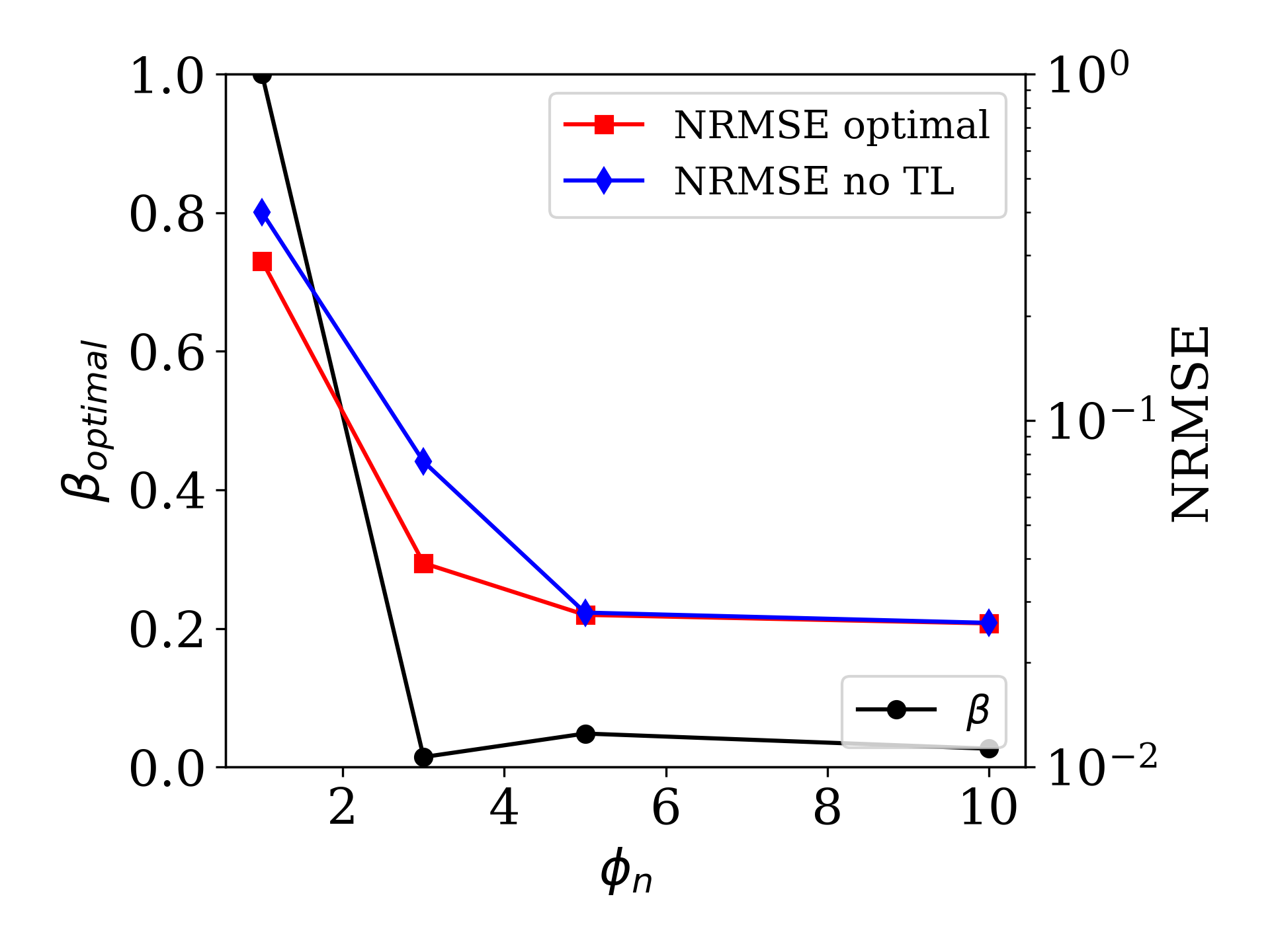}
  \caption{Transfer of the last 4 PC modes}
\end{subfigure}
\caption{Optimal $\beta$ as a function of number of equivalence ratios for different unburned temperatures for the target task for the ammonia-H$_2$ mechanism.}
\label{fig:betaOpt_partial}
\end{figure}

A test was performed for the ammonia-H$_2$ fuel where transfer learning was performed for only the last autoencoder weights (PC modes). It should be noted that for the ammonia-H$_2$ data set, a total of 14 modes were retained to capture 99.9 \% of the variance. Two configurations have been tested: one where the last 4 modes out of 14 were transferred, and another one where the last 10 modes were transferred. In both cases, the low rank modes were initialized with an uninformative prior with a flat distribution. Figure \ref{fig:betaOpt_partial} presents the results for the transfer of knowledge using all PC modes (Fig.~\ref{fig:betaOpt_partial}a) and the tests of transferring only part of the PC modes. The variation of $\beta_{optimal}$ with the number of equivalence ratios in the target task behaves similarly to the previous analysis where a larger transfer of knowledge is performed with decreasing $\phi_n$. However, a larger NRMSE is observed as the number of PC modes used for knowledge transfer is reduced. The results indicate that the transfer of knowledge in the low-rank modes is important for the good performance of the methodology presented in this study.

\begin{figure}[h!]
    \centering
    \includegraphics[width=3.5in,keepaspectratio=true]{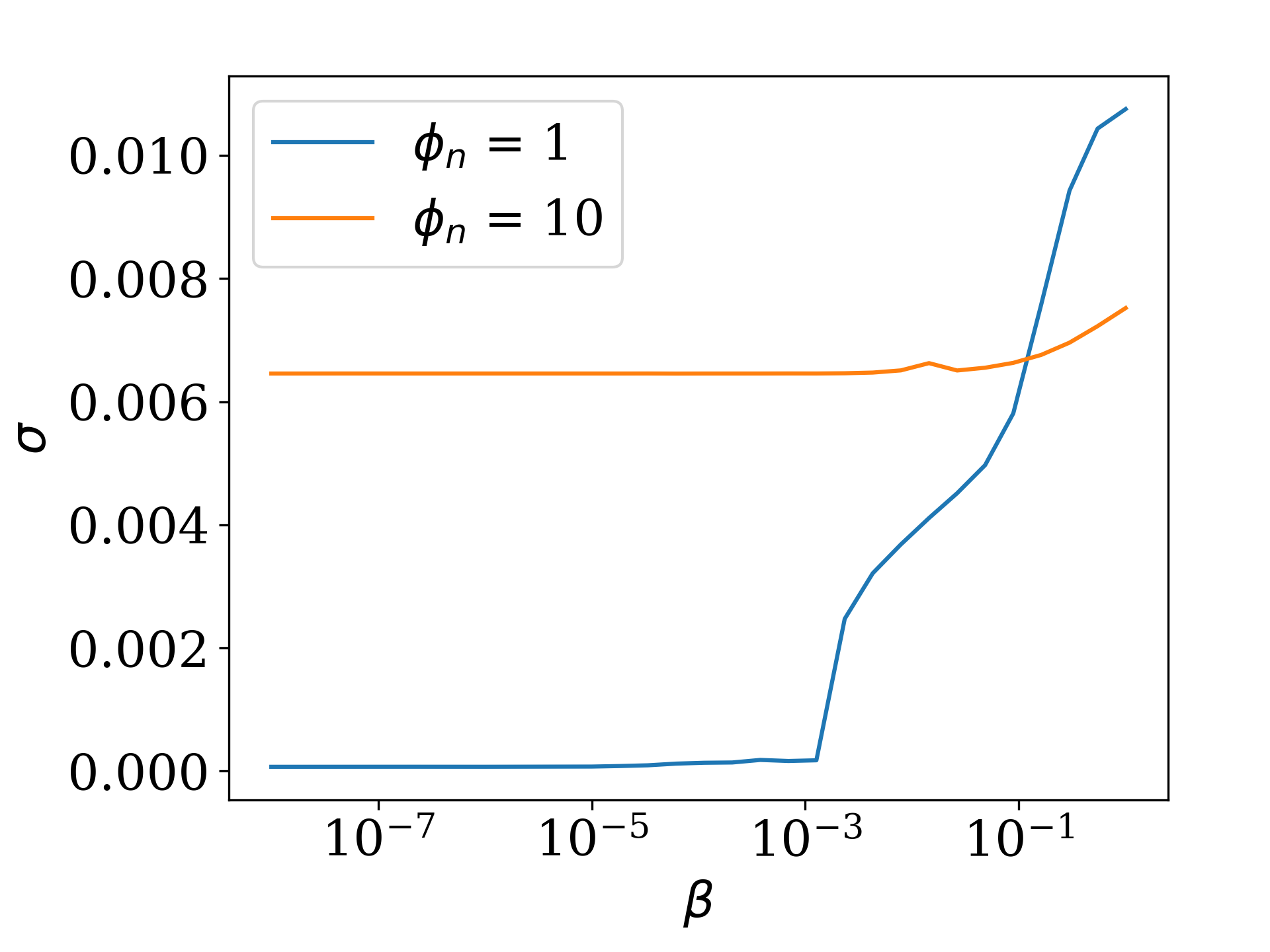}
    \caption{Variance inferred by the Bayesian neural network for H$_2$-ammonia flames at $T_u$ = 300.}
    \label{fig:sigma_evolution_Tu_300.0}
\end{figure}

The objective function used in this work is assumed to be linear with a measurement error denoted by $\sigma$ in Eq.~(\ref{ML_model}). The $\sigma$ value corresponds to the data that cannot be captured by the retained number of dimensions employed in the dimensionality reduction. In other words, the data outside of the lower dimensional manifold is predicted by autoencoders. The expected behavior for this metric is to correlate with the reconstruction error of the data since $\sigma$ is essentially capturing the limitation in the objective function. Figure \ref{fig:sigma_evolution_Tu_300.0} presents the variation of $\sigma$ with respect to the transfer learning parameter $\beta$ for different amounts of data for the target task. Starting with $\phi_n = 1$, the $\sigma$ value remains close to zero for low transfer learning ($\beta \to 0$) due to the model adaptation to accurately capture the single flame used in the target task. However, as transfer learning increases the autoencoder weights are modified by a more informative prior which increases the captured error for the target task data. In the case of $\phi_n = 10$, a larger $\sigma$ is associated with a greater amount of data to be captured by the number of dimensions employed in the dimensionality reduction. A small increase in $\sigma$ for $\phi_n = 10$ is related to negative transfer learning identified in Fig \ref{fig:l2norm_phi}. Therefore, transfer learning through Bayesian autoencoder weights corresponds to a special case for $\sigma$ which in order to gain model generalizability to minimize the reconstruction error for the different conditions, i.e. different equivalence ratios, there is a performance reduction in the target task training reflected in the predicted $\sigma$ for $\phi_n = 1$ as $\beta \to 1$. 

The Bayesian method models each of the autoencoder weights (latent variables) as a Gaussian distribution with a mean and a variance. A lower variance in the model indicates a higher confidence of the model in correctly predicting the latent space, where in the limit of a perfect reconstruction of the features the variance should approach zero. Figure \ref{fig:weights_scale} presents the contour map for the variance of the latent variables as a function of the number of scalars used and the number of retained dimensions. The results correspond to the latent space for the target task with $\beta = \beta_{optimal}$ using different amounts of data: $\phi_n = 1$ and $\phi_n = 10$. A general increase in the variance of the latent variables can be observed as the data sparsity is increased for each fuel. The model produces the correct behavior as it tends to be less certain with $\phi_n = 1$ compared to $\phi_n = 10$ despite keeping a constant NRMSE for the reconstructed data (Fig.~\ref{fig:betaOpt_phi}). Equation \ref{toy_problem}, originally developed for a simplified problem, can be used to gain a further understanding of the effects of the predicted variance for the autoencoder weights in the transfer learning methodology presented here. The variance both in the source and target tasks controls the amount of diffusion of knowledge between tasks. A larger variance in the autoencoder weights for the target task ($\sigma_t$) results in a lower contribution of the target task mean weights ($\mu_t$) in the mean posterior evaluation during the transfer learning optimization process. The variance in the autoencoder weights for the source task ($\sigma_s$) dictates the contribution from the source task mean weights ($\mu_s$) and also regulates the tempering parameter $\beta$ for different weights. Generally, the model allows for a larger diffusion of knowledge from the source task for weights that present a lower variance.

As previously mentioned, there is no enforcement in the autoencoders to rank the weights in the latent space by the amount of variance captured in the data. The initialization strategy with the eigenvectors obtained with PCA provides a ranked latent space close to the optimal solution. Figure \ref{fig:weights_scale} shows a clear reduction in the predicted variance for the lower ranks in the latent space across all three chemical mechanisms and sparsity levels. The behavior helps in the compatibility of the lower dimensional manifolds between the source and target tasks to perform transfer learning since the weights are sorted by their importance in capturing the variance in the data.

\subsection{Comparison with the deterministic transfer learning approach}
Lastly, the performance of the proposed transfer learning approach is compared with that of the state-of-the-art deterministic transfer learning strategy, which was introduced by \cite{Li18, Li20}. In a deterministic machine learning model, Li et al. \cite{Li18, Li20} recently suggested that the knowledge obtained from the source task can be partially-transferred to the target task by adjusting the magnitude of $\alpha$ in the penalty term, which is defined by:

\begin{equation}
  \Omega(\textbf{\textit{w}}) = \frac{\alpha}{2} || \textbf{\textit{w}} - {\textbf{\textit{w}}^{0}} ||_2^2
  \label{TL3}
\end{equation}
where $\Omega(\textbf{\textit{w}})$ represents the regularization term, $\alpha$ the regularization parameter that controls the strength of the penalty, $\textbf{\textit{w}}$ and ${\textbf{\textit{w}}^{0}}$ the parameter vectors that are adapted to the target task and that are obtained from the source task, respectively, and $||\cdot||_p$ the \textit{p}-norm of a vector. Then, a fine-tuning process for the target task proceeds with a trade-off between the data-fitting term (e.g., mean squared error loss function) and regularization term, $\Omega(\textbf{\textit{w}})$. The knowledge obtained from the source task is fully transferred to the target task (i.e., parameter freezing) as $\alpha$ reaches infinite, whereas the machine learning model in the target task is trained from scratch (i.e., without transfer learning) as $\alpha$ becomes near zero. Note that the role of $\alpha$ in Eq.~\ref{TL3} is somewhat analogous to that of $\beta$ introduced in the present study, but the distinct feature of the probabilistic transfer learning method proposed in the present study is that the regularization induced is through a Bayesian mechanism. The proposed strategy relies on capturing and propagating prior parameter knowledge (obtained from source task) in the form of prior probability density functions which not only encode point estimates ${\textbf{\textit{w}}^{0}}$ but also uncertainties and correlation information. The hypothesis is that this additional information (relating to parametric uncertainties) allows for a more informative transfer of knowledge and ultimately provides a more robust mechanism of knowledge transfer.

To reasonably compare the performances between the deterministic and the present probabilistic transfer learning methods, a set of deterministic linear autoencoder models is also prepared by adopting the same dataset used in this study. For the training of deterministic linear autoencoder models, the training dataset in the source task corresponds to the 15 different 1-D freely-propagating flames with $T_u$ of 400 K and various equivalence ratios ranging from 0.5 to 2.0 (i.e., $\phi_n$ = 15), as described in Sec.~\ref{equi_sparse}. The datasets in the target tasks consist of several different 1-D flames with $T_u$ of 300 K and the same range of equivalence ratios with the source task, but relatively low $\phi_n$ (i.e., $\phi_n \leq 15$). Three different pairs of the source and target tasks are prepared with the use of each of three different chemical kinetic mechanisms, namely ammonia-H$_2$, C1, and TPRF mechanisms. The dimensionality of the low-dimensional manifold for the cases with ammonia-H$_2$, C1, and TPRF mechanisms are 14, 30, and 33, respectively. 

For both source and target tasks, 80\% of the training dataset is used for training and the remaining 20\% of the training dataset is used for validation set to to evaluate the performance of the model and prevent overfitting. The parameters of the autoencoder are initialized by applying PCA to the dataset, and subsequently, the parameters of the deterministic linear autoencoder models are optimized by using the mean squared error (MSE) loss function. The Adam optimizer \cite{Adam} is used for stochastic optimization. After training the autoencoder in the source task, the knowledge from the source task is utilized for the training of the target task by varying the magnitudes of $\alpha$ in Eq.~(\ref{TL3}). The best achievable value of NRMSE (i.e., minimum NRMSE) from a series of training results with various $\alpha$ is then evaluated at a given $\phi_n$. 

\begin{figure}[h!]
    \centering
    \includegraphics[width=3in,keepaspectratio=true]{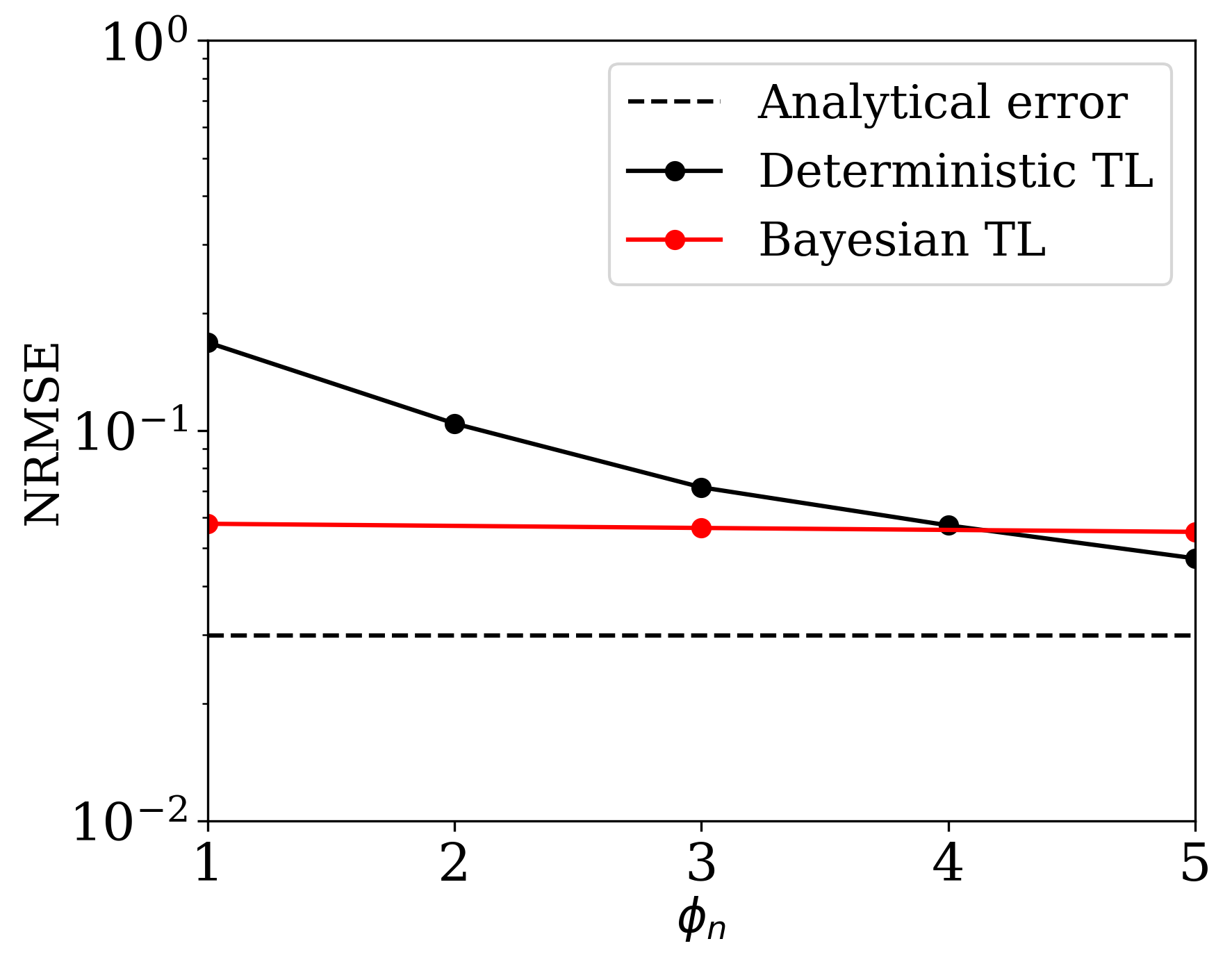}
    \caption{Comparison between different transfer learning strategies.}
    \label{fig:comparison}
\end{figure}

Figure~\ref{fig:comparison} presents the variations in the best achievable NRMSE in the target tasks as a function of $\phi_n$ by applying the transfer learning method proposed in the present study or the deterministic transfer learning model. The probabilistic method provides a reconstruction error (NRMSE) approximately 3 times lower compared to the deterministic method when $\phi_n = 1$ or it requires 4 times less data to achieve the same NRMSE, demonstrating the advantage of applying the probabilistic transfer learning framework. 

\section{Conclusions}
 The present study proposes a solution for the data sparsity problem related to turbulent combustion models based on data-driven low-dimensional manifold transport. A novel transfer learning methodology based on Bayesian neural networks is presented, where information is transferred between tasks in terms of lower dimensional manifold obtained during the dimensionality reduction. A source task with abundant data is trained with the same Bayesian neural network setting in order to obtain a posterior distribution that is used as a prior in the target task. The amount of information provided to the target task through the prior distribution is controlled by the parameter $\beta$ that dictates how informative the prior distribution is in estimating the posterior distribution, i.e. the latent space.  The neural network methodology is applied to two different simplified tests using laminar freely-propagating flames: the first one where data sparsity in the target task is imposed by spatially sampling the one-dimensional flame solution; and another one where data sparsity in the target task is imposed by selecting a given number of flames with different equivalence ratios from a larger data set with the same oxidizer temperature.

 The spatial sparsity test used for the source task a fully resolved freely-propagating flame solution at a stoichiometric equivalence ratio and 300 K for the oxidizer temperature. The target task used samples of a freely-propagating flame solution at an equivalence ratio 1.2 and 300 K for the oxidizer temperature. The results showed that transfer learning can reduce the deviations in the reconstruction of the scalars in physical space. The methodology was able to reduce the overall reconstruction error in a data-sparse setting when compared to a no-transfer learning strategy. A moderate transfer of knowledge was obtained for the spatial sparsity test, with $\beta_{optimal}$ ranging between $10^{-2}$ and $10^{-4}$. Nonetheless, an order of magnitude reduction in the NRMSE was obtained for the larger mechanism used (TPRF 98 species). The low value for the $\beta_{optimal}$ in this test is attributed to the relatively small similarity between the source and target tasks. Negative transfer learning was observed for the condition with the tempering parameter $\beta$ was large for abundant data in the target task.

 The second test with data sparsity in equivalence ratio space showed a large reduction in the NRMSE with transfer learning in a large data sparsity setting, i.e. when only 1 to 3 equivalence ratios out of 15 are selected for the target task. The transfer learning strategy was also able to keep the reconstruction error constant as the data sparsity increased. A 10-fold reduction in data requirement was achieved in order to have the same NRMSE observed for the abundant case scenario with the TPRF mechanism. The reduction in data requirement is of a factor of 5 for the smaller chemical mechanisms. The transfer learning strategy was also able to identify the similarity between tasks and increase/decrease the tempering parameter $\beta$ for similar/dissimilar conditions. However, tests indicate that the diffusion of knowledge has to be performed for all the latent space in order to obtain a good performance. 

 Comparisons between the Bayesian transfer learning approach and a state-of-the-art deterministic transfer learning approach were performed using the most complex chemical mechanism tested. The results showed that the Bayesian method can outperform the deterministic approach in a large data sparsity scenario. A 3 times reduction in the NRMSE was observed when only one equivalence ratio solution out of 15 was used for the target task. The probabilistic method also requires 4 times less data to obtain the same reconstruction error observed for the deterministic model.

A transfer learning strategy based on the diffusion of knowledge in lower dimensional space has proven to be effective in reducing the reconstruction error at conditions with data sparsity. The methodology can be directly applied to combustion reactors, as presented in Ref. \cite{Zhang2022}, and CFD. Future work involves the addition of the tempering parameter $\beta$ in the optimization process and the application of the new strategy \textit{a posteriori} in a DNS context.

\section*{Acknowledgements}\label{Acknowledgements}
This work was supported by the Laboratory Directed Research and Development program at Sandia National Laboratories (Project 222361), a multimission laboratory managed and operated by National Technology and Engineering Solutions of Sandia LLC, a wholly-owned subsidiary of Honeywell International Inc. for the U.S. Department of Energy’s National Nuclear Security Administration under contract DE-NA0003525. This report describes objective technical results and analysis. Any subjective views or opinions that might be expressed in the report do not necessarily represent the views of the U.S. Department of Energy or the United States Government.

\bibliographystyle{elsarticle-num}
\bibliography{references}

\end{document}